\documentclass[lettersize,journal]{IEEEtran}
\usepackage{cite}
\usepackage{amsmath,amssymb,amsfonts}
\usepackage{graphicx}
\usepackage{textcomp}
\usepackage{dblfloatfix}
\usepackage[caption=false,font=normalsize,labelfont=sf,textfont=sf]{subfig}
\usepackage{xcolor}

\usepackage{bm}

\newcounter{defcounter}
\DeclareMathOperator{\st}{s.t.}
\DeclareMathOperator{\col}{col}

\usepackage{cases}
\usepackage{array}

\usepackage{empheq}

\usepackage{dsfont}

\newcommand{\mybold}{\text{\usefont{U}{bbold}{m}{n}1}}
\MakeRobust{\mybold}

\usepackage{bm}

\begin{document}
%
\title{Targeted Algorithmic Purpose-Driven Cyber Attacks in Distributed Multi-Agent Optimization}

\author{Mahan Fakouri Fard$^{\dagger}$ and Mingxi Liu$^{\dagger}$, ~\IEEEmembership{Member,~IEEE}
\thanks{$^{\dagger}$M. Fakouri Fard and M. Liu are with the Department of Electrical and Computer Engineering at the University of Utah, 50 S Central Campus Drive, Salt Lake City, UT, 84112, USA {\tt{\{mahan.fakourifard, mingxi.liu\}@utah.edu}}.}}

\markboth{Manuscript submitted to IEEE Transactions on Industrial Cyber-Physical Systems}%
{Shell \MakeLowercase{\textit{et al.}}: A Sample Article Using IEEEtran.cls for IEEE Journals}


\maketitle
\IEEEpeerreviewmaketitle
\begin{abstract}
 Distributed multi-agent optimization (DMAO) enables the scalable control and coordination of a large population of edge resources in complex multi-agent environments. Despite its great scalability, DMAO is prone to cyber attacks as it relies on frequent peer-to-peer communications that are vulnerable to malicious data injection and alteration. Existing cybersecurity research mainly focuses on \emph{broad-spectrum} attacks that aim to jeopardize the overall environment but fail to sustainably achieve specific or targeted objectives. This paper develops a class of novel strategic purpose-driven algorithmic attacks that are launched by participating agents and interface with DMAO to achieve self-interested attacking purposes. Theoretical foundations, in both primal and dual senses, are established for these attack vectors with and without stealthy features.  Simulations on electric vehicle charging control validate the efficacy of the proposed algorithmic attacks and show the impacts of such attacks on the power distribution network.
\end{abstract}


\section{Introduction}\label{sec I introduction}
\IEEEPARstart{A}{s} digital systems grow in scale and complexity, traditional centralized control frameworks are increasingly challenged by the vast amount of data and diversity of network‑edge resources. Cyber‑physical applications, e.g., modern power systems, are integrating large numbers of distributed energy resources, electric vehicles (EV), and other edge devices that demand efficient coordination. Distributed multi-agent optimization (DMAO) algorithms have thus emerged as a pivotal advancement in cyber-physical systems \cite{Xu_2018_iot,sanislav_2017_multiagent}. By distributing and/or parallelizing the decision-making process across multiple agents, DMAO frameworks enable rapid response and optimized resource allocation within complex networked environments. However, the very features that confer scalability also introduce new challenges in ensuring system security.

A generic multi-agent optimization problem for $v$ agents can be formulated as
\begin{equation}
\min_{\bm{x}\in\mathcal{X}}  {\sum_{i=1}^{v}f_i(\bm{x}_i)+F(\bm{x})}, \label{DMAO problem}
\end{equation}
where $\mathcal{X}=\{\bm{x}|\bm{h}(\bm{x})\in \mathbb{H}, \bm{x}_{i} \in \mathbb{X}_{i},\forall i=1,2, \ldots, v \}$, 
$\bm{x} \in \mathbb{R}^{\eta}$ is the collection of all decision variables $\bm{x}_i$'s,  $f_i(\cdot):\mathbb{R}^{\eta_i}\mapsto\mathbb{R}$ is the competitive local objective of agent $i$, $F(\cdot):\mathbb{R}^{\eta}\mapsto\mathbb{R}$ is the coupled cooperative network-level objective,  $\mathbb{X}_i$ is the local constraint set, $\bm{h}(\cdot):\mathbb{R}^{\eta}\mapsto\mathbb{R}^{\iota}$ is the network constraint mapping that normally couples all $\bm{x}_i$'s \cite{huo_2022_twofacet}, and $\mathbb{H}$ denotes the network operation constraint set. 

Popular classes of DMAO algorithms for solving \eqref{DMAO problem} include the alternating direction method of multipliers (ADMM) \cite{boyd2011admm}, average consensus \cite{Nedic_TAC_2010}, particle swarm methods \cite{kennedy_1995_particle}, and projection-based methods \cite{zhang_2015_distributed}.  
The DMAO algorithms have shown outstanding performance in controlling edge resources and revolutionizing power grid services. For example, in \cite{ADMMEV_Zhou_2021, ADMMexchange_khaki_2019}, ADMM-based distributed control schemes were developed to coordinate the charging of a large population of EVs.  \cite{DecentralizedEvSPDS_Liu} and \cite{huo2022two} introduced shrunken-primal-dual subgradient (SPDS) to construct distributed EV charging control frameworks. In \cite{Zheng_2023_concensus}, an average-consensus-based \textsc{On}/\textsc{Off} control of air conditioner clusters was developed for continuous balancing reserves in smart grids.  In general, the DMAO algorithms follow certain updating and communication protocols to iteratively update the decision variables and reach convergence, where the update at the $k$th iteration assumes 
\begin{equation}
\label{general update}
\bm{x}_i^{(k+1)}=U(\bm{x}_i^{(k)},\mathcal{G}_i(\{\bm{x}_{j}^{(k)}\}_{j\in\mathcal{N}_i})).
\end{equation}Herein,  $U(\cdot):\mathbb{R}^{\eta_i} \mapsto \mathbb{R}^{\eta_i}$ is a local update mapping that maps agent $i$’s current state and neighbor information to its next state, $\mathcal{G}_i(\cdot)$ specifies how agent $i$ and/or the system operator collects and processes the states of agent $i$'s neighbors ($\mathcal{N}_i$) before passing them to the update function. The specific forms of $U(\cdot)$ and $\mathcal{G}_i(\cdot)$ are algorithm‑dependent. 


Systems running by DMAO algorithms inherently rely on frequent peer-to-peer communications, exposing them to multiple potential vulnerabilities that adversaries can exploit. Traditional cybersecurity research in this domain has predominantly focused on \emph{broad-spectrum} attacks such as denial-of-service (DoS) \cite{Cheng_2022_dos}, man-in-the-middle (MITM) attacks \cite{vu_2024_mitm}, false data injection (FDI) \cite{dehbozorgi_2025_fdia}, and replay attacks \cite{towardefficient_cao_2017}. These attacks typically induce large-scale disruptions, characterized by duration and impact, as they degrade system performance and destabilize network control, thereby compromising the stability and reliability of critical infrastructures. Although such attacks are highly effective in creating observable disturbances \cite{vu_2024_mitm}, their overt impact often triggers rapid detection by system operators who rely on real-time monitoring and anomaly detection algorithms \cite{ zheng_2024_fdiaCounter}. Consequently, while these conventional attack strategies can severely impair system operations, their brute-force nature and high visibility limit their sustainability as long-term threats.
In contrast, \emph{strategic purpose-driven} algorithmic cyber attacks \cite{Mahan_2023_cyber} represent a subtler yet potentially more pernicious threat vector. In these scenarios, participating agents -- integral components of a DMAO framework -- intentionally and strategically manipulate their and/or other agents' iterative updates to achieve targeted, self-serving objectives without noticeably compromising the overall convergence of the algorithm. Such carefully calibrated manipulations are engineered to yield asymmetric benefits. In \cite{voltageregulation_liu_2022}, an attacker secured preferential resource allocation while maintaining a facade of normalcy that evades standard detection protocols \cite{CybersecADMM_Munsing_2018}. The sophistication of these attacks lies in their gradual introduction of bias, where minuscule alterations in each iteration cumulatively steer the system toward outcomes that are favorable to the attacker. This insidious strategy exploits the inherent trust relationships and the distributed nature of the system, thereby challenging the efficacy of conventional anomaly detection. 

Existing research on this front, however, is sparse, and only few truly considered targeted or self-beneficial attacks. Munsing and Moura \cite{CybersecADMM_Munsing_2018} examined distortions, noise injection, and coupling‑constraint manipulation in ADMM and developed detection and mitigation schemes, demonstrating that distributed optimization is vulnerable to such attacks. Sundaram and Gharesifard \cite{sundaram_2018_distributed} showed that consensus-based methods are vulnerable to nodes deviating from prescribed updates and designed a resilient algorithm that guarantees the non-adversarial agents converge to the convex hull of local minimizers under graph-robustness conditions. Yemini \emph{et al.} \cite{yemini_2025_resilient} developed a stochastic‑trust framework for multi‑agent cyber‑physical systems, which, however, only targets generic adversarial disturbances rather than strategic self‑interested manipulation. Despite these advances, existing studies primarily target broad-spectrum adversaries, e.g., Byzantine and faulty nodes, whose goal is to degrade convergence or consensus. They lack the capability to explicitly address targeted or self-interested, stealthy manipulations that steer DMAO iterations towards an attacker's private objective while preserving the manner of nominal operation. In our previous work \cite{Mahan_2023_cyber}, we preliminarily explored such attacks to enable the agents to strategically inject data into the DMAO iterations, aiming to achieve self-beneficial outcomes without convergence disruption or noticeable operation constraint violations. Though this work represents an important first step in pioneering purpose-driven algorithmic attacks, it does not encompass the full scope of the field and lacks a rigorous theoretical foundation.

This paper aims to transcend the conventional understanding of cyber attacks within DMAO systems, filling the gap between generic disruption and strategic manipulation by broadening our vision to encompass attacks engineered for particular intents. In particular, this paper formulates and analyzes purpose-driven algorithmic attacks and discusses targeted offenses tailored to DMAO's update and communication structure. The contribution of this paper is two-fold:
\begin{enumerate}
    \item Theoretical foundations are established for the strategic purpose-driven algorithmic attacks within DMAOs. 
    \item A novel dual attacking strategy is developed to coordinate manipulations across agents and steer system outcomes while preserving convergence, revealing a new undercover threat vector within DMAOs.
\end{enumerate}

This paper is organized as follows. Section \ref{sec II problem formulation} presents the problem formulation, the necessary preliminaries, and a leading EV charging control example. Section \ref{sec III main} formalizes strategic purpose‑driven algorithmic cyber attacks,  analyzes their fundamental properties and impacts, and provides rigorous proofs. Section \ref{sec IV results} validates the theoretical foundations on convergence and stealthiness through a comprehensive case study in the context of EV charging control. Finally, Section \ref{sec V conclusion} concludes the paper and outlines avenues for future work.


\section{Problem Formulation and Preliminaries} \label{sec II problem formulation}



To better illustrate the concept of strategic purpose-driven algorithmic attacks, we adopt an EV charging control problem within the distribution network as a vertical-driven example to formulate a DMAO problem as in \eqref{DMAO problem}, where each EV is an agent and the distribution network is the environment.

\subsection{Distribution network and EV charging model}
As the focus of this paper is on the investigation of cyber attacks, we intend to simplify the physics of both the distribution network and EV charging. To this end, the linear model of a single-phase distribution network with $n$ buses, at time $t$, can be written as \cite{baran1989optimal} 
\begin{equation} \label{voltage}
\bm{V}(t)=\bm{V}_{0}-2 \bm{R} \bm{P}(t)-2 \bm{X} \bm{Q}(t),
\end{equation}
where $\bm{V}(t) \in \mathbb{R}^{n}$ encompasses the squared voltage magnitudes for all buses while $\bm{V}_0 = V_0^2 \bm{1}_n \in \mathbb{R}^{n}$ signifies the constant voltage magnitude vector at the slack bus, with $V_0$ denoting the voltage magnitude at the feeder head. Furthermore, $\bm{P}(t) \in \mathbb{R}^{n}$ and $\bm{Q}(t) \in \mathbb{R}^{n}$ represent the real and reactive power consumption across all buses, respectively.  $\bm{R}\in \mathbb{R}^{n \times n}$ and $\bm{X}\in \mathbb{R}^{n \times n}$ are the resistance and reactance of the line adjacency matrices, respectively. Detailed descriptions of the LinDisFlow model can be found in \cite{farivar2013equilibrium}.

The power consumption at each bus is simplified to include a known baseline load and the controllable aggregated EV charging load. Assuming the EVs only consume real power, at bus $l$, we have\begin{equation} \label{power}
    p_l(t)=p_{l,b}(t)+p_{l,EV}(t)\end{equation} and $q_l(t)=q_{l,b}(t)$, where $p_{l,b}(t)$, $q_{l,b}(t)$, and $p_{l,EV}(t)$ denote the real baseline power, reactive baseline power, and EV charging power, respectively. Let $\bm{V}_b(t)$ denote the  voltage drop caused by the baseline load, \eqref{voltage} can be rewritten as
\begin{equation}
    \bm{V}(t)=\bm{V}_0 - \bm{V}_b(t) - 2 \bm{R}\bm{p}_{EV}(t),
\end{equation} where $\bm{p}_{EV}(t)=[p_{1,EV}(t) ~ \cdots~  p_{n,EV}(t)]^{\mathsf{T}}$. Let $s$ be the total number of EVs and $s_l$ be the number of EVs connected to bus $l$, we have $p_{l,EV}(t)= \sum_{\hat{l}=1}^{s_{l}}\bar{P}_{l,\hat{l}}c_{l,\hat{l}}(t),$
where $c_{l,\hat{l}}(t)$ and $\bar{P}_{l,\hat{l}}$ are the percentage charging rate and maximum charging power of the $\hat{l}$th EV connected to bus $l$. For simplicity and clarity, we re-index the subscripts of $c_{l,\hat{l}}$ and $\bar{P}_{l,\hat{l}}$ with $i=1,\ldots,s$ by following the ascending orders of $l$ and $\hat{l}$. By defining $\bm{G}=\oplus_{l=1}^n \bm{1}_{s_l}^{\mathsf{T}} \in \mathbb{R}^{n\times s}$ and $\bar{\bm{P}}=\oplus_{i=1}^s \bar{P}_i \in \mathbb{R}^{s\times s}$, we have
\begin{equation}
    \bm{V}(t)=\bm{V}_0 - \bm{V}_b(t) - 2 \bm{R}\bm{G}\bar{\bm{P}}\bm{C}(t),
\end{equation} where $\bm{1}_{s_l}^{\mathsf{T}}$ is the charging power aggregation vector, $\bm{C}(t)=[c_1(t) ~ c_2(t) ~ \cdots ~ c_s(t)]^{\mathsf{T}} \in \mathbb{R}^s$, and $\oplus$ denotes matrix direct sum. Further let $\bm{D}\in\mathbb{R}^{n\times s}$ denote $- 2 \bm{R}\bm{G}\bar{\bm{P}}$, $\bm{y}_d(t)$ denote $\bm{V}_0 - \bm{V}_b(t)$ and $\bm{y}(t)$ denote $\bm{V}(t)$, we have
\begin{equation}
    \bm{y}(t)= \bm{y}_d(t) + \bm{D}\bm{C}(t).
\end{equation}

Let $SOC_{i,ini}$ and $SOC_{i,des}$ denote the initial and the desired state of charge (SOC), respectively, and $\hat{E}_i$ denote the battery capacity of $i$th EV. Then the total energy required by the $i$th EV is $E_{i,req}=\hat{E}_i(SOC_{i,des}-SOC_{i,ini}).$
\subsection{Valley filling -- a cooperative multi-agent optimization}

Valley filling via EV charging control leverages the aggregated EV charging power overnight to minimize the variance of the total load profile of a distribution network, thereby assisting utilities in lowering operational costs. Let $T$ be the valley filling period, then the charging profile of the $i$th EV is denoted as $ \bm{c}_i=[c_i(t)~c_i(t+1)~\cdots~c_i(t+T-1)]^{\mathsf{T}} \in \mathbb{R}^T$ and $\bm{\mathcal{C}} = [\bm{c}_1^{\mathsf{T}}  \cdots \bm{c}_s^{\mathsf{T}} ]^{\mathsf{T}} \in \mathbb{R}^{sT}$ is the collection of all EVs' charging profiles. The valley-filling problem can then be formulated as\begin{subequations} \label{main problem}
    \begin{align}
\min_{\bm{\mathcal{C}}} & ~ \mathcal{F}(\bm{\mathcal{C}})=\frac{1}{2}\left\|\bm{P}_b+\sum_{i=1}^{s}\bar{P_i}\bm{c}_i\right\|_2^2 \label{3a}\\
    \st & ~ \bm{c}_i\in\mathbb{C}_i, ~ \forall i \in 1,2,...,s, \label{3b}\\
    & ~ \bm{\mathcal{Y}}_b-\sum_{i=1}^n\bm{\mathcal{D}}_i\bm{c}_i\leq \bm{0},
    \label{3c}
    \end{align}
\end{subequations}
where $\bm{P}_b = [P_b(t)~P_b(t+1)~~\cdots ~~ P_b(t+T-1)]^{\mathsf{T}}\in\mathbb{R}^T$ is the aggregated baseline load profile of the entire distribution network.  $\mathbb{C}_i$ ensures the $i$th EV's charging request is fulfilled by the end of the valley filling, which takes the form of
\begin{equation} \label{primal constraints}
   \mathbb{C}_i:=\{\bm{c}_i| \bm{0} \leq \bm{c}_i \leq \bm{1}, E_{i,req}-\bm{\hat{B}}_{i,l}\bm{c}_i=0\},
    \end{equation}
    where $\bm{\hat{B}}_{i,l}= \bm{1}_s^{\mathsf{T}}\bm{B}_{i,l}$, $\bm{B}_{i,l}=[\bm{B}_{i,c}~\bm{B}_{i,c}~\cdots~\bm{B}_{i,c}]\in \mathbb{R}^{s\times T}$, $\bm{B}_{i,c}\in \mathbb{R}^s$ denotes the the $i$th column of  $\bm{B}=\oplus_{i=1}^s B_i$, $B_i=-\eta_i\bm{\Delta} t\bar{P_i}$, $\eta_i$ is the charging efficiency, and $\Delta t$ is the sampling time. Eqn. \eqref{3c} ensures all nodal voltage magnitudes stay above the lower bound, where  $\bm{\mathcal{Y}}_b$ denotes $\underline{v}^2\bm{V_0}-\bm{\mathcal{Y}}_{d}$, $\bm{\mathcal{Y}}_d=[\bm{y}_d(t)~\bm{y}_d(t+1)~\cdots~\bm{y}_d(t+T-1)]^{\mathsf{T}}\in \mathbb{R}^{nT}$, $\underline{v}$ is the bus voltage magnitude lower bound, $\bm{\mathcal{D}_i}=D_i\oplus D_i\dots\oplus D_i\in\mathbb{R}^{sT\times T}$ denotes the mapping between EV charging power and the nodal voltage magnitudes, and $\bm{D}=[D_1~D_2~\cdots~D_s]$. For clarity, \eqref{3b} adopts the simplest EV charging formulation, enforcing battery energy and network voltage constraints only. Practical implementations often include additional network/device limits and objectives, which are compatible with the same modeling framework.

\subsection{Distributed EV charging control scheme}
We adopt SPDS \cite{DecentralizedEvSPDS_Liu} as an example DMAO algorithm to construct a distributed paradigm of solving \eqref{main problem}. With the relaxed Lagrangian
of problem \eqref{main problem} defined as 
\begin{equation}
    \mathcal{L}(\bm{\mathcal{C}},\bm{\mu}) = \mathcal{F}(\bm{\mathcal{C}}) + \bm{\mu}^{\mathsf{T}} \big(\bm{\mathcal{Y}}_b-\sum_{i=1}^n\bm{\mathcal{D}}_i\bm{c}_i\big), 
\end{equation}
each EV updates its primal variable at iteration $k$ by following 
\begin{equation} \label{primal update}
     \bm{c}_i^{(k+1)} {=} \Pi_{\mathbb{C}_i}\left(\frac{1}{\tau_\mathcal{C}}\Pi_{\mathbb{C}_i}\left(\tau_\mathcal{C}\bm{c}_i^{(k)}{-}\alpha_{i,k}\nabla_{\bm{c}_i}\mathcal{L}(\bm{\mathcal{C}}^{(k)},\bm{\mu}^{(k)})\right)\right),
    \end{equation}
    where $\Pi$ represents the Euclidean projection \cite{DecentralizedEvSPDS_Liu}, $\tau_\mathcal{C}$ is the primal shrinking parameter and $\alpha_{i,k}$ is the primal update step size. Similarly, the dual variable updates by following
    \begin{equation} \label{dual update}
     \bm{\mu}^{(k+1)} = \Pi_{\mathbb{D}}\bigg(\frac{1}{\tau_\mu}\Pi_{\mathbb{D}}\big(\tau_\mu \bm{\mu}^{(k)}+\beta_{k}\triangledown_{\mu}\mathcal{L}(\bm{\mathcal{C}}^{(k)},\bm{\mu}^{(k)})\big)\bigg),
\end{equation}
where $\bm{\mu}\in \mathbb{R}^{nT}$ is the dual variable associated with \eqref{3c}, $\tau_\mu$ is the dual shrinking parameter, and $\beta_{i,k}$ is the dual update step size. $\mathbb{D}$ is the dual optimal set and is constructed to guarantee the convergence.
By implementing SPDS in EV charging control, individual EVs only need to share their own $\bm{c}_i^{(k)}$ with the system operator. The system operator computes the Lagrange gradient and $\bm{\mu}^{(k)}$ and broadcasts them to all EVs. This process will continue until the tolerance $\|\bm{\mathcal{C}}^{(k+1)}-\bm{\mathcal{C}}^{(k)}\|_2$ drops below a threshold $\epsilon$. The convergence and optimality proofs of SPDS can be referred to \cite{DecentralizedEvSPDS_Liu}. 


\section{Purpose-Driven Algorithmic Cyber Attacks}\label{sec III main}
In this section, we use the EV charging control problem to establish the theoretical foundation of the purpose-driven algorithmic attacks.  We will introduce multiple attack scenarios where attackers only manipulate their own data, i.e., primal attack scenarios, and where attackers manipulate other EVs' data to gain personal advantages, i.e., dual attack scenarios.

 

\subsection{Targeted algorithmic primal attack vectors} \label{sec_self_attack}

Let the $i$th EV be an attacker seeking a self-interest goal represented by $g_i(\bm{c}_i)$. We present Theorem 1 that establishes the equivalence between the distributed malicious local updates and the centralized under-attack problem represented by\begin{subequations} \label{attack problem}
    \begin{align}
    &\min_{\bm{\mathcal{C}}}~\mathcal{F}(\bm{\mathcal{C}}) +\omega_1g_i(\bm{c}_i)  \label{8a}\\
    &  \st ~~\eqref{3b},~\eqref{3c},
    \end{align}
\end{subequations}where $\omega_1>0$ denotes the power of the self-interest attack. 

\noindent \textbf{Theorem 1} \cite{Mahan_2023_cyber}: An agent participating in a distributed projected gradient descent algorithm can deviate the converged solution of \eqref{DMAO problem} towards its target goal, represented by a convex $g_i(\bm{c}_i)$ with bounded subgradients, without affecting the algorithm convergence by only locally altering its primal update direction by $\omega_1\nabla g_i(\bm{c}_i^{(k)})$, where $\omega_1>0$ is the attacking power. \hfill $\blacksquare$

Theorem 1 shows that an attacker can pursue self-interest goals while following the DMAO algorithm and not interfering with algorithm convergence. These self-interest goals can be related to the attacker's charging profile, like the timing and speed/rate of charging, while they can also affect other EVs, like damaging their battery health. Before introducing the next theorem, we first present the following assumptions to ensure the existence of minimizers and allow the usage of subgradient‑based optimality conditions.

\noindent\textbf{Assumption 1:} Set $\mathbb{S}\subseteq\mathbb{R}^n$ is nonempty, closed and convex, and it holds Slater's condition for problem \eqref{attack problem2}. \hfill $\square$

\noindent\textbf{Assumption 2:} Mapping $g_i:\mathbb{S}_i\mapsto\mathbb{R}$ is proper, $L_{g_{i}}$-smooth, closed and convex, and with bounded subgradients. \hfill $\square$

\noindent\textbf{Assumption 3:} Mapping $\mathcal{F}:\mathbb{S}\mapsto\mathbb{R}$ is continuously differentiable and $m$-strongly convex. \hfill $\square$


Define the general multi-attacker problem as\begin{subequations} \label{attack problem2}
    \begin{align}
    &\min_{\bm{\mathcal{C}}}~\mathcal{F}(\bm{\mathcal{C}}) + G(\bm{\mathcal{C}})  \label{}\\
    &  \st ~~\eqref{3b},~\eqref{3c},
    \end{align}
\end{subequations} where $G(\bm{\mathcal{C}})=\sum_{i=1}^s \omega_i g_i(\bm{c}_i)$ is the aggregated attack function and $\omega_i$ represents the local attacking power. Let $\mathbb{S}$ be the feasible region and let  $\bm{\mathcal{C}}^\star \in \mathbb{S}$ and $\bm{\mathcal{C}}_A^\star \in \mathbb{S}$ be the unique solutions of \eqref{main problem} and \eqref{attack problem2}, respectively. Let $\boldsymbol{\sigma}^\star$ be a member of $\partial G(\bm{\mathcal{C}}^\star)$,  $\psi(\bm{\mathcal{C}}^\star)$ denote the tangent cone of $\mathbb{S}$ at $\bm{\mathcal{C}}^\star$, and $\Pi_{\psi(\bm{\mathcal{C}}^\star)}$ be the Euclidean projector onto $\psi(\bm{\mathcal{C}}^\star)$.

\noindent\textbf{Theorem 2:}  Under {Assumptions 1}-{3}, the deviation between the attack-free and under-attack optimizers is bounded by
    \begin{equation} \label{thorem21}
      \left\|\bm{\mathcal{C}}_A^\star - \bm{\mathcal{C}}^\star\right\|_2 
      \le \frac{1}{m}\left\|\Pi_{\psi(\bm{\mathcal{C}}^\star)}\boldsymbol{\sigma}^\star\right\|_2
      \le \frac{1}{m}B,
 \end{equation}where $B=\sqrt{\sum_{i=1}^n\omega_iL_{g_i}^2}$. In addition, the deviation between the attack-free and under-attack optimal values is bounded  by \begin{equation} \label{thorem22}
      0\le \mathcal{F}(\bm{\mathcal{C}}_A^\star)-\mathcal{F}(\bm{\mathcal{C}}^\star)
      \le \frac{1}{2m}\left\|\Pi_{\psi(\bm{\mathcal{C}}^\star)}\bm{\sigma}^\star\right\|_2^2.
    \end{equation} If, in addition, $\nabla \mathcal{F}$ is Lipschitz with constant $L_{\mathcal{F}}$, then 
    \begin{equation}\label{thorem22b}
      \mathcal{F}(\bm{\mathcal{C}}_A^\star)-\mathcal{F}(\bm{\mathcal{C}}^\star)\le \tfrac{L_{\mathcal{F}}}{2}\left\|\bm{\mathcal{C}}_A^\star-\bm{\mathcal{C}}^\star\right\|_2^2.
      \end{equation}Finally, writing $\bm{\mathcal{C}}_A^\star=\col(\bm{c}_{1,A}^\star,\dots,\bm{c}_{s,A}^\star)$ and $\boldsymbol{\sigma}^\star=\col(\boldsymbol{\sigma}_1^\star,\dots,\boldsymbol{\sigma}_s^\star)$ and letting $\boldsymbol{\psi}_i$ be the $i$-th block of $\psi(\bm{\mathcal{C}}^\star)$, one has
    \begin{equation} \label{thorem23}
      \left\|\bm{\mathcal{C}}_{i,A}^\star - \bm{\mathcal{C}}_i^\star\right\|_2 \le \frac{1}{m}\left\|\Pi_{\psi_i}\bm{\sigma}_i^\star\right\|_2 \le \frac{1}{m}L_{g_i}.
    \end{equation} \hfill $\blacksquare$

\noindent\textbf{Proof:}
Because $\mathcal{F}$ is $m$-strongly convex and each $g_i$ is convex, the function $\mathcal{F}+ G$ is also $m$-strongly convex.  Thus, problems in \eqref{main problem} and \eqref{attack problem2} both admit unique solutions 
To prove \eqref{thorem21}, let $N_{\mathbb{S}}(\cdot)$ denote the normal cone of the feasible set $\mathbb{S}$. By first–order optimality, there exist $\bm{v}^\star\in N_{\mathbb{S}}(\bm{\mathcal{C}}^\star)$, $\bm{v}_A\in N_{\mathbb{S}}(\bm{\mathcal{C}}^\star_A)$, and $\bm{\sigma}_A\in\partial G(\bm{\mathcal{C}}^\star_A)$ such that
\begin{align}
\nabla_{\bm{\mathcal{C}}}\mathcal{F}(\bm{\mathcal{C}}^\star) + \bm{v}^\star &= \bm{0}, \\
\nabla_{\bm{\mathcal{C}}}\mathcal{F}(\bm{\mathcal{C}}^\star_A) + \bm{\sigma}_A + \bm{v}_A&=\bm{0}.
\end{align}
Let $\bm{\Delta}\triangleq\bm{\mathcal{C}}^\star_A-\bm{\mathcal{C}}^\star$, it leads to
\begin{align}
\bm{\Delta}^{\mathsf{T}}\left(\nabla_{\bm{\mathcal{C}}}\mathcal{F}(\bm{\mathcal{C}}^\star_A)-\nabla_{\bm{\mathcal{C}}}\mathcal{F}(\bm{\mathcal{C}}^\star)\right)+\bm{\Delta}^{\mathsf{T}}\left(\bm{v}_A - \bm{v}^\star\right) & \nonumber \\
+\bm{\Delta}^{\mathsf{T}}\left(\bm{\sigma}_A - \bm{\sigma}^\star\right) &= - \bm{\Delta}^{\mathsf{T}} \bm{\sigma}^\star.
\end{align}
Since $\nabla_{\bm{\mathcal{C}}}\mathcal{F}$ is $m$-strongly monotone, the first term on the left is at least $m\|\bm{\Delta}\|_2^2$.  Monotonicity of the normal cone $N_{\mathbb{S}}$ and monotonicity of the subdifferential $\partial G$ imply
$ \bm{\Delta}^{\mathsf{T}}(\bm{v}_A - \bm{v}^\star)\ge 0$ and 
$\bm{\Delta}^{\mathsf{T}}(\bm{\sigma}_A - \bm{\sigma}^\star)\ge 0$.
Hence
\begin{equation}\label{22}
m\|\bm{\Delta}\|^2 \le -\bm{\Delta}^{\mathsf{T}} \bm{\sigma}^\star.
\end{equation}
Because $\bm{\mathcal{C}}^\star_A$ minimises $\mathcal{F}+ G$, one has $G(\bm{\mathcal{C}}^\star_A) \le G(\bm{\mathcal{C}}^\star)$. Further, convexity of $G$ implies $\bm{\Delta}^{\mathsf{T}} \bm{\sigma}^\star \le 0$. Decompose $\bm{\sigma}^\star$ onto the tangent and normal planes at $\bm{\mathcal{C}}^\star$ as $\Pi_{\psi(\bm{\mathcal{C}}^\star)}\bm{\sigma}^\star$ and $\Pi_{N(\bm{\mathcal{C}}^\star)}\bm{\sigma}^\star$, respectively. Since $\bm{\Delta}^{\mathsf{T}} \Pi_{N(\bm{\mathcal{C}}^\star)}\bm{\sigma}^\star\le 0$, we have
\begin{equation}
\bm{\Delta}^{\mathsf{T}} \bm{\sigma}^\star
\le
\bm{\Delta}^{\mathsf{T}} \Pi_{N(\bm{\mathcal{C}}^\star)}\bm{\sigma}^\star
\le
\|\Pi_{\psi(\bm{\mathcal{C}}^\star)}\bm{\sigma}^\star\|_2\|\bm{\Delta}\|_2.
\end{equation}Substituting \eqref{22} into the inequality above and canceling $\|\bm{\Delta}\|_2$ yields $\|\bm{\Delta}\|_2 \le \frac{1}{m}\|\Pi_{\psi(\bm{\mathcal{C}}^\star)}\bm{\sigma}^\star\|_2.$ By boundedness of the subgradients of $g_i$, we have $\|\Pi_{\psi(\bm{\mathcal{C}}^\star)}\bm{\sigma}^\star\|\le \|\bm{\sigma}^\star\|\le B$, leading to the results in \eqref{thorem21}.

In the following, we prove \eqref{thorem22} and \eqref{thorem23}. From the optimality of $\bm{\mathcal{C}}^\star_A$, one has $
\mathcal{F}(\bm{\mathcal{C}}^\star_A) - \mathcal{F}(\bm{\mathcal{C}}^\star)
\le
G(\bm{\mathcal{C}}^\star) - G(\bm{\mathcal{C}}^\star_A).$ Since the convexity of $G$ implies $G(\bm{\mathcal{C}}^\star_A)\ge G(\bm{\mathcal{C}}^\star) +  \bm{\Delta}^{\bm{\mathsf{T}}}\bm{\sigma}^\star$, we have the distance bound $\mathcal{F}(\bm{\mathcal{C}}^\star_A) - \mathcal{F}(\bm{\mathcal{C}}^\star)
\le
-\bm{\Delta}^{\bm{\mathsf{T}}}\bm{\sigma}^\star.$
Using the same decomposition of $\bm{\sigma}^\star$ as in proving \eqref{thorem21} and the Cauchy–Schwarz inequality  yields
\begin{equation}
\left|\bm{\Delta}^{\bm{\mathsf{T}}}\bm{\sigma}^\star\right|
\le
\|\Pi_{\psi(\bm{\mathcal{C}}^\star)}\bm{\sigma}^\star\|_2\|\bm{\Delta}\|_2.
\end{equation}
Combining with the distance bound proved above and applying the arithmetic–geometric inequality gives \eqref{thorem22}. Further, if $\nabla\mathcal{F}$ has Lipschitz constant $L_{\mathcal{F}}$, the standard smoothness inequality yields the alternative bound in \eqref{thorem22b}. 

Because $g_i$ is Lipschitz, there exists $L_{g_i}\ge0$ such that any subgradient $\bm{\sigma}_i\in\partial g_i(\bm{c}_i)$ satisfies $\|\bm{\sigma}_i\|_2\le L_{g_i}$. To prove \eqref{thorem23}, we project the distance bound onto the $\boldsymbol{\psi}_i$ and use $\|\bm{\sigma}_i^\star\|_2\le L_{g_i}$, leading to \eqref{thorem23}. \hfill$\Box$

In the following, we introduce a lemma to relax {Assumption 3} to further generalize {Theorem 2}, where, without strong convexity, the weak‐sharp‐minima property provides an error–bound condition that still controls the solution deviations.

\noindent \textbf{Lemma 1:}  Under {Assumptions 1} and {2}, given a convex $\mathcal{F}$ and the problem in \eqref{main problem} admitting a weak sharp minimum with modulus $\alpha>0$,  then for any $\bm{\mathcal{C}}_A^\star\in S^\star_A$, there exists $\bm{\mathcal{C}}^\star\in S^\star$ and $\bm{\sigma}^\star\in\partial G(\bm{\mathcal{C}}^\star)$ such that
\begin{equation}\label{Lemma_1_ineq}
  \mathrm{dist}\left(\bm{\mathcal{C}}_A^\star,S^\star\right)
  \le
  \frac{1}{\alpha}\left\|\Pi_{\psi(\bm{\mathcal{C}}^\star)}\bm{\sigma}^\star\right\|_2
  \le
  \frac{1}{\alpha}B,
\end{equation}
where $S^\star$ and $S^\star_A$ are the solution sets of \eqref{main problem} and \eqref{attack problem2}, respectively, and $\mathrm{dist}(\bm{x},S)\triangleq\inf_{\bm{y}\in S}\|\bm{x}-\bm{y}\|_2$.   \hfill $\blacksquare$

\noindent \textbf{Proof:} Since $\bm{\mathcal{C}}_A^\star$ minimises $\mathcal{F}+ G$, one has
\(\mathcal{F}(\bm{\mathcal{C}}_A^\star)+ G(\bm{\mathcal{C}}_A^\star)\le\mathcal{F}(\bm{\mathcal{C}}^\star)+ G(\bm{\mathcal{C}}^\star)\) for any $\bm{\mathcal{C}}^\star\in S^\star$.  Weak sharpness of $\mathcal{F}$ implies $\mathcal{F}(\bm{\mathcal{C}}_A^\star)\ge\mathcal{F}^\star + \alpha\mathrm{dist}(\bm{\mathcal{C}}_A^\star,S^\star)$, and convexity of $G$ gives $G(\bm{\mathcal{C}}_A^\star)\ge G(\bm{\mathcal{C}}^\star)+(\bm{\mathcal{C}}_A^\star-\bm{\mathcal{C}}^\star)^{\bm{\mathsf{T}}}\bm{\sigma}^\star$.  Combining these inequalities yields 
\(\alpha\mathrm{dist}(\bm{\mathcal{C}}_A^\star,S^\star)\le -(\bm{\mathcal{C}}_A^\star-\bm{\mathcal{C}}^\star)^{\bm{\mathsf{T}}}\bm{\sigma}^\star\).  Since the normal component of $\bm{\sigma}^\star$ at $\bm{\mathcal{C}}^\star$ has non–positive inner product with any feasible direction, we have $\alpha\mathrm{dist}(\bm{\mathcal{C}}_A^\star,S^\star)\le\|\Pi_{\psi(\bm{\mathcal{C}}^\star)}\bm{\sigma}^\star\|_2\|\bm{\mathcal{C}}_A^\star-\bm{\mathcal{C}}^\star\|_2$.  Because $\mathrm{dist}(\bm{\mathcal{C}}_A^\star,S^\star)\le\|\bm{\mathcal{C}}_A^\star-\bm{\mathcal{C}}^\star\|_2$ and $\|\Pi_{\psi(\bm{\mathcal{C}}^\star)}\bm{\sigma}^\star\|_2\le B$, the bound in \eqref{Lemma_1_ineq} is proved. \hfill $\square$


Based on {Theorem 1}, our preliminary work \cite{Mahan_2023_cyber} introduced the smooth-charging attack, the rush-charging attack, and their stealthy variants. To avoid repetition, we refer the readers to \cite{Mahan_2023_cyber} for detailed discussions on the smooth-charging attack, where attackers seek flattened charging profiles, and the stealthy attacks, where attackers seek to minimize deviations in the converged results while still achieving their objectives.

In the following, we generalize the rush-charging attack, whose key idea is to set preferred charging times by assigning different weights to the attacker’s decision variables. This leads to the following extended cases.

\subsubsection{Time tuning attack}
An attacking EV may want itself to be fully charged earlier or select its own preferred charging periods. This can be achieved by injecting $2\omega_1 \bm{A}^{\mathsf{T}} \bm{A}\bm{c}_i^{(k)}$ at each primal update iteration, where $\bm{A}\in \mathbb{R}^{T\times T}$ is a diagonal reshaping matrix with its elements defined as
\begin{equation}
A_{\hat{t},\hat{t}}=\left\{ \begin{array}{ll}
m,&~ \text{if } \hat{t} \in \Theta, \\
M,&~ \text{otherwsie.}
\end{array}\right.
\end{equation}
Herein, $0<m \ll M$ and $\Theta\subseteq\{1,2,\dots,T\}$ is the set of the attacker's preferred charging time indices. Note that $\Theta$ can contain any time slots, even if they are not consecutive. According to {Theorem 1}, this is equivalent to adding $\omega_1\|\bm{A}\bm{c}_i\|_2^2$ to the objective function of problem \eqref{main problem}.

\subsubsection{Battery damage attack}
Unlike other presented attack vectors that pursue personal benefits, the battery damage attack will harm the attacking EVs. In other words, the attacking EVs are actually the victims of an external attacker who plans to damage and demobilize the affected EVs by forcing them to charge in a high-oscillatory pattern. To this end, it needs to inject $2\omega_1 \hat{\bm{A}}^{\mathsf{T}} \hat{\bm{A}}\bm{c}_i^{(k)}$ at each primal update iteration, where $\hat{\bm{A}}\in \mathbb{R}^{T\times T}$ is a diagonal reshaping matrix, similar to that in the previous scenario, with each diagonal element defined as
\begin{equation}
\hat{A}_{\hat{t},\hat{t}}=\left\{ \begin{array}{ll}
m,&~ \text{if } \frac{\hat{t}}{t_f} \in \mathbb{I} \\
M,&~ \text{otherwise,}
\end{array}\right.
\end{equation}    
where $0<m \ll M$, $t_f$ denotes the attacker's desired oscillation frequency of charge, and $\mathbb{I}$ is a set of integers. According to {Theorem 1}, this is equivalent to adding $\omega_1\|\hat{\bm{A}}\bm{c}_i\|_2^2$ to the objective function of problem \eqref{main problem}.
Entries in $\hat{\bm{A}}$ with smaller values will force the corresponding elements in $\bm{c}_i$ to be maximized, and \emph{vice versa}, to achieve the battery damage goal.

\subsection{Dual targeted algorithmic cyber-attack}\label{sec dual attack}
In a targeted algorithmic attack, remaining undetected is crucial for the attackers because achieving the attack goals relies on persistently attacking until the algorithm converges. In \cite{Mahan_2023_cyber}, we have discussed the enhanced stealthiness by suppressing the objective difference between the under-attack and attack-free cases. However, all the pre-designed data is injected by the attackers into their own updates. This fact increases the possibility of the attackers getting caught if the system operators get suspicious and analyze the data from the agents' communication channels while utilizing anomaly-based \cite{zhang_2006_anomaly} or machine learning-based detection methods \cite{batchu_2021_generalized}. To overcome this issue and enhance the stealthiness of the attack, we propose a novel concept -- \emph{dual} cyber-attack, using which the attacker injects data into the receiving channels of other innocent agents to achieve similar attacking performance as it injects data into its own updating procedure. By doing so, the attacker keeps its own communication channel uncontaminated, but exploits others to perform the attack. 

To this end, the attacker should exploit some system-level governing constraints to design its specific dual attack vectors. In our design, we let the attacker manipulate the data in the nodal load balance equation \eqref{power}, leading innocent agents to receive falsified data related to \eqref{power} and converge to a solution that benefits the attacker. Specifically, the attacker first exploits the innocent agents' communication channels to send corrupted, deceptive data to the system operator, thereby falsifying the nodal load balance equation. The attacker then forces the system operator to send deceptive data to the innocent agents, maintaining the falsified nodal load balance. In essence, the attacker falsifies the nodal load balance constraint data in a way that forces the algorithm to solve a centralized problem with a virtually additional $g_i$ term in the objective. Because the constraints are unchanged and strong duality still holds, according to Assumption 1, this strategy forces the system to converge to the same solution that would result if EV $i$ had directly added $g_i$ to its objective in the primal update. Thus, the dual attack can realize the same optimal $\bm{\mathcal{C}}^\star$ and objective value as the primal attack, under the same chosen goal represented by $g_i$. In the following, we formalize this by asserting that the dual attack problem (with the crafted $g_i$) yields the same converged solution as the primal attack case.  

\noindent \textbf{Assumption 4:} The attackers have full knowledge of the operating algorithms, and they can wiretap and inject data into the I/O of the system operator during the attack. \hfill $\square$

Let the $i$th EV be the attacker, we have the following dual attack theorem.


\noindent \textbf{Theorem 3:} Under {Assumptions 1}-{4}, there exists a data injection mapping $\Phi_i:\mathbb{S}\times\mathbb{R}^n\mapsto\mathbb{R}^n$ into the dual update of \eqref{main problem} such that the resulting dual attack solution $\bm{\tilde{\mathcal{C}}}^\star$ satisfies $\bm{\tilde{c_i}}^\star=\bm{\hat{c_i}}^\star$, where $\bm{\hat{\mathcal{C}}}^\star$ is the solution of primal attack \eqref{attack problem}. \hfill $\blacksquare$

\noindent \textbf{Proof:} Let $R(\bm{\mathcal{C}})=\bm0$ denote the general form of one of the system coupling constraints that the system operator enforces in any projected gradient descent based DMAO algorithm. Then the relaxed Lagrangian of \eqref{main problem} can be rewritten as
\begin{equation}
L(\bm{\mathcal{C}},\bm{\lambda})=\mathcal{F}(\bm{\mathcal{C}})+\bm{\lambda}^{\bm{\mathsf{T}}}R(\bm{\mathcal{C}})+\bm{\mu}^{\bm{\mathsf{T}}}Q(\bm{\mathcal{C}}),
\end{equation} where $Q(\bm{\mathcal{C}})$ denotes \eqref{3b} with $R(\bm{\mathcal{C}})$ extracted.

In the primal attack problem \eqref{attack problem} and according to {Theorem 1}, $\omega_1g_i$ is added to the Lagrangian, while in the dual attack problem, the attacker injects a falsification function $\Phi_i$ into the operator's dual update. Equivalently, the operator behaves as if its Lagrangian were
\begin{equation}
L(\bm{\mathcal{C}},\bm{\lambda})=\mathcal{F}(\bm{\mathcal{C}})+\bm{\lambda}^{\mathsf{T}} \bigl(R(\bm{\mathcal{C}})+\Phi_i(\bm{\mathcal{C,\bm{\lambda}}})\bigr)+\bm{\mu}^{\bm{\mathsf{T}}}Q(\bm{\mathcal{C}}).
\end{equation} Thus, at a stationary point $(\bm{\mathcal{C}},\bm{\lambda})$, we have 
\begin{equation}\label{stationary}
\nabla \mathcal{F}(\bm{\mathcal{C}}) + \nabla R(\bm{\mathcal{C}})^{\mathsf{T}} \bm{\lambda} + \nabla_{\bm{\mathcal{C}}}\big(\bm{\lambda}^{\mathsf{T}} \Phi_i(\bm{\mathcal{C}},\bm{\lambda})\big) + \bm{v}\ =\ \bm 0,
\end{equation}where $\bm{\mu}^{\bm{\mathsf{T}}}Q(\bm{\mathcal{C}})=\bm{0}$, $R(\bm{\mathcal{C}})=\bm 0$, $\bm{v}\in N_{\mathbb{S}}(\bm{\mathcal{C}})$, and $N_{\mathbb{S}}(\bm{\mathcal{C}})$ is the Eculidean normal cone at $\bm{\mathcal C}\in\mathbb{S}$. Under the primal attack \eqref{attack problem}, the stationarity is instead \begin{equation}\label{stationary2}
\nabla \mathcal{F}(\bm{\mathcal{C}}) + \omega_1\nabla g_i(\bm{\mathcal{C}}) + \nabla R(\bm{\mathcal{C}})^{\mathsf{T}} \bm{\lambda} + \bm{v} =\ \bm 0.\end{equation} Hence, to match the primal attack at the optimizer, it suffices to make the dual-side injected term equal the primal $g_i$ term, i.e., $\nabla_{\bm{\mathcal{C}}}\big(\bm{\lambda}^{\mathsf{T}} \Phi_i(\bm{\mathcal{C}},\bm{\lambda})\big)=\omega_1\nabla g_i(\bm{\mathcal{C}})$. Consequently, we define \begin{equation}
    \Phi_i(\bm{\mathcal{C}},\bm{\lambda}) \ \triangleq\ \frac{\omega_1}{\|\bm{\lambda}\|^2}g_i(\bm{\mathcal{C}})\bm{\lambda} 
\end{equation} $\forall \bm{\lambda} \neq \bm{0}$ and set $\Phi_i(\bm{\mathcal{C}},\bm0)=\bm0$. Let $(\hat{\bm{\mathcal{C}}}^\star,\hat{\bm{\lambda}}^\star,\hat{\bm{\mu}}^\star)$ be the KKT pair of \eqref{attack problem2}. With the $\Phi_i$ constructed as above, the dual attack stationarity \eqref{stationary2} at $(\hat{\bm{\mathcal{C}}}^\star,\hat{\bm{\lambda}}^\star)$ becomes \begin{equation}\label{stationary3}
\nabla \mathcal{F}(\hat{\bm{\mathcal{C}}}^\star) + \omega_1\nabla g_i(\hat{\bm{\mathcal{C}}}^\star) + \nabla R(\hat{\bm{\mathcal{C}}}^\star)^{\mathsf{T}} \hat{\bm{\lambda}}^\star + \bm{v} =\ \bm 0,\end{equation} with the same feasibility $R(\hat{\bm{\mathcal{C}}}^\star)=\bm 0$ and $\bm{v}\in N_{\mathbb{S}}(\hat{\bm{\mathcal{C}}}^\star)$. Thus $(\hat{\bm{\mathcal{C}}}^\star,\hat{\bm{\lambda}}^\star)$ also satisfies the KKT conditions of the dual attack problem. Since $\mathcal{F}$ is $m$-strongly convex on $\mathbb{S}$, both problems have a unique primal optimizer, hence $\bm{\tilde{c}}_i^\star=\bm{\hat{c}}_i^\star$. \hfill $\square$


{Assumption 4} provides the attackers with full access and knowledge of the system operator. In what follows, we present a Lemma that denies the attacker's access to the dual variables to construct $\Phi_i$, which is a more practical case. We show how the attacker can implement the dual attack with limited knowledge access by manipulating the nodal load balance equality and using the stealthy technique.

\noindent \textbf{Lemma 2:} Under {Assumption 1-3}, there exists a mapping $g_i(\bm{\mathcal{C}})$ such that the dual attack problem can be represented as\begin{subequations} \label{dual attack problem}
    \begin{align}
    &\min_{\bm{\mathcal{C}}}~\mathcal{F}(\bm{\mathcal{C}}) +\omega_3g_i(\bm{\mathcal{C}})  \label{8a}\\
    & ~ \st ~\eqref{3b},~\eqref{3c},
    \end{align}
\end{subequations}
where the dual attack solution $\bm{\tilde{\mathcal{C}}}^\star$ satisfies $\bm{\tilde{c_i}}^\star\approx\bm{\hat{c_i}}^\star$. Herein, $\bm{\hat{\mathcal{C}}}^\star$ is the solution of primal attack \eqref{attack problem}, $\nabla_{\bm{c}_i}g_i(\bm{\mathcal{C}})=\bm{0},~ \forall i\in\mathbb{A}$, and $\mathbb{A}$ is the set of attacker's index. \hfill $\blacksquare$ 

\noindent \textbf{Proof:} Assume the $i$th EV is connected to bus $l$ with $s_l$ other EVs. The power consumption at bus $l$ is $ p_l(t)=p_{l,b}(t)+p_{l,EV}(t)$, where $p_{l,EV}(t)=\bar{P}_{l,i}c_{l,i}(t)+\sum_{\hat{l}=1}^{s_{l}}\bar{P}_{l,\hat{l}}c_{l,\hat{l}}(t).$ Further, we can write $c_{l,i}(t)=\bar{p}_{l,i}(t)-\sum_{\hat{l}=1}^{s_{l}}\bar{P}_{l,\hat{l}i}c_{l,\hat{l}}(t),$
where $\bar{p}_{l,i}(t)=\left(p_l(t)-p_{l,b}(t)\right)/{\bar{P}_{l,i}}$ and $\bar{P}_{l,\hat{l}i}={\bar{P}_{l,\hat{l}}}/{\bar{P}_{l,i}}$. Extending $\bar{p}_{l,i}(t)$ for all time slots and removing the attacker's bus index can lead to
\begin{equation}
\bm{c}_i=\bm{\bar{\mathcal{P}}}_i-\sum_{\hat{l}=1}^{s_{l}}\bm{\bar{\mathcal{P}}_{\hat{l}i}}\bm{{c}}_{\hat{l}}.\end{equation}
Suppose the attacker could manipulate the information of the real baseline power and total power injection to bus $l$, i.e., $p_l(t)-p_{l,b}(t)$, in a way that $\bm{\bar{\mathcal{P}}}_i$ becomes a constant vector over all iterations. Therefore, the function $g_i(\bm{c}_i)$ represented in {Theorem 1} can be represented as $g_i(\bm{c}_{\hat{l}}|\hat{l}=1,\dots,s_l)$. Therefore, following {Theorem 1}, as long as $g_i(\bm{c}_{\hat{l}}|\hat{l}=1,\dots,s_l)$ is convex and $\omega_3>0$, the algorithm convergence is guaranteed. 

In contrast to the primal attack and the dual attack under {Assumption 4}, in this scenario, the attackers' challenge is to find a proper $\bm{\bar{\mathcal{P}}}_i$ and maintain the tampered power balance equation $\bm{c}_i=\bm{\bar{\mathcal{P}}}_i-\sum_{\hat{l}=1}^{s_{l}}\bm{\bar{\mathcal{P}}_{\hat{l}i}}\bm{{c}}_{\hat{l}}$ instead of finding a proper $g_i(\bm{c}_i)$ for gaining personal benefits or constructing $\Phi_i$. To this end, the attacker has to estimate the value of $\bm{\bar{\mathcal{P}}}_i$ and $\sum_{\hat{l}=1}^{s_{l}}\bm{\bar{\mathcal{P}}_{\hat{l}i}}\bm{{c}}_{\hat{l}}$ at the optimal solution, to form its attack vector via implementing the same technique as in the stealthy for-purpose algorithmic primal attack. Specifically, at the start of the algorithm, the $i$th EV hibernates by allowing the algorithm to run normally without launching any attack. As the iteration goes, once $\hat{\epsilon}^{(\ell)}$ drops below $\epsilon_s$ in the $\ell$th iteration, the $i$th EV regards the algorithm ``converged'' and wiretaps other EVs' communication channels to obtain $\bm{\bar{\mathcal{P}}}_i$ and $\sum_{\hat{l}=1}^{s_{l}}\bm{\bar{\mathcal{P}}_{\hat{l}i}}\bm{{c}}_{\hat{l}}$ which will be used as an approximation of $\bm{\mathcal{C}}^\star$. At any iteration $k$ after the $(\ell+1)$th iteration, the $i$th EV launches the dual attack by injecting
$\bm{\bar{\mathcal{P}}}_i-\sum_{\hat{l}=1}^{s_{l}}\bm{\bar{\mathcal{P}}_{\hat{l}i}}\bm{{c}}_{\hat{l}}$ through other EVs' primal update to achieve a self-interest goal. Note that {Lemma 2} does not guarantee that the solutions of dual and prime attack scenarios are equivalent. However, based on \textbf{Theorem 2}, the deviations in optimal solution and values are bounded. \hfill $\square$ 

\section{Simulation Results}\label{sec IV results}
We demonstrate the effectiveness of the proposed attack scenarios through simulations of controlling 500 EVs connected to a simplified single-phase IEEE 13-bus test feeder. See \cite{Mahan_2023_cyber} for detailed simulation configurations. Since the simulation results of stealthy and non-stealthy smooth-charging attacks and rush-charging attacks were already reported in our preliminary work \cite{Mahan_2023_cyber}, herein we omit those results but focus on presenting new results on battery-damage attacks and dual attacks

\subsection{Primal attack scenarios}
\subsubsection{Attack-free scenario}

\begin{figure*}[t]
  \centering
  (a){\includegraphics[width=0.30\textwidth]{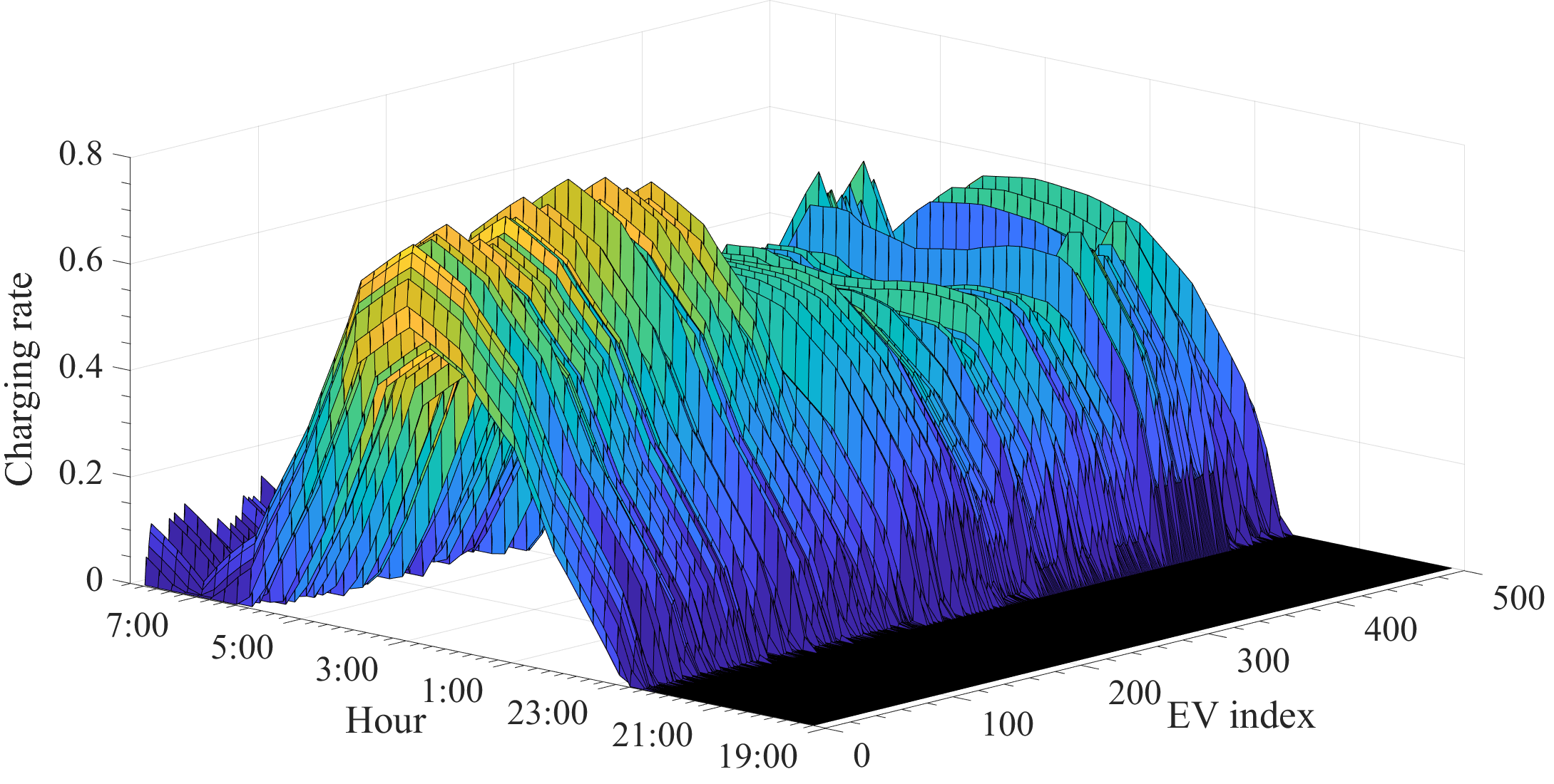}\label{no attack u}}\hfill
  (c){\includegraphics[width=0.30\textwidth]{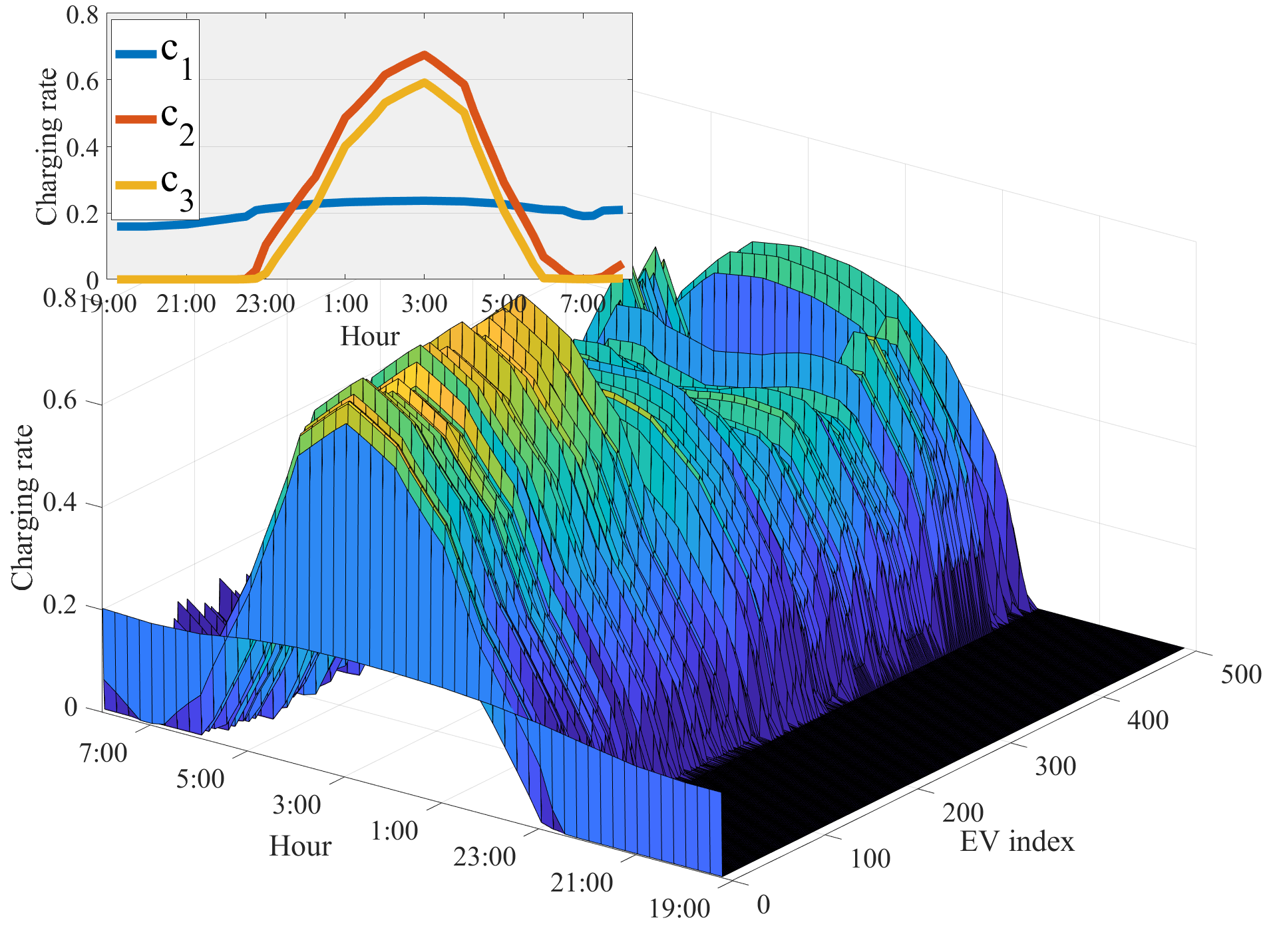}\label{primal attack u}}\hfill
  (e){\includegraphics[width=0.30\textwidth]{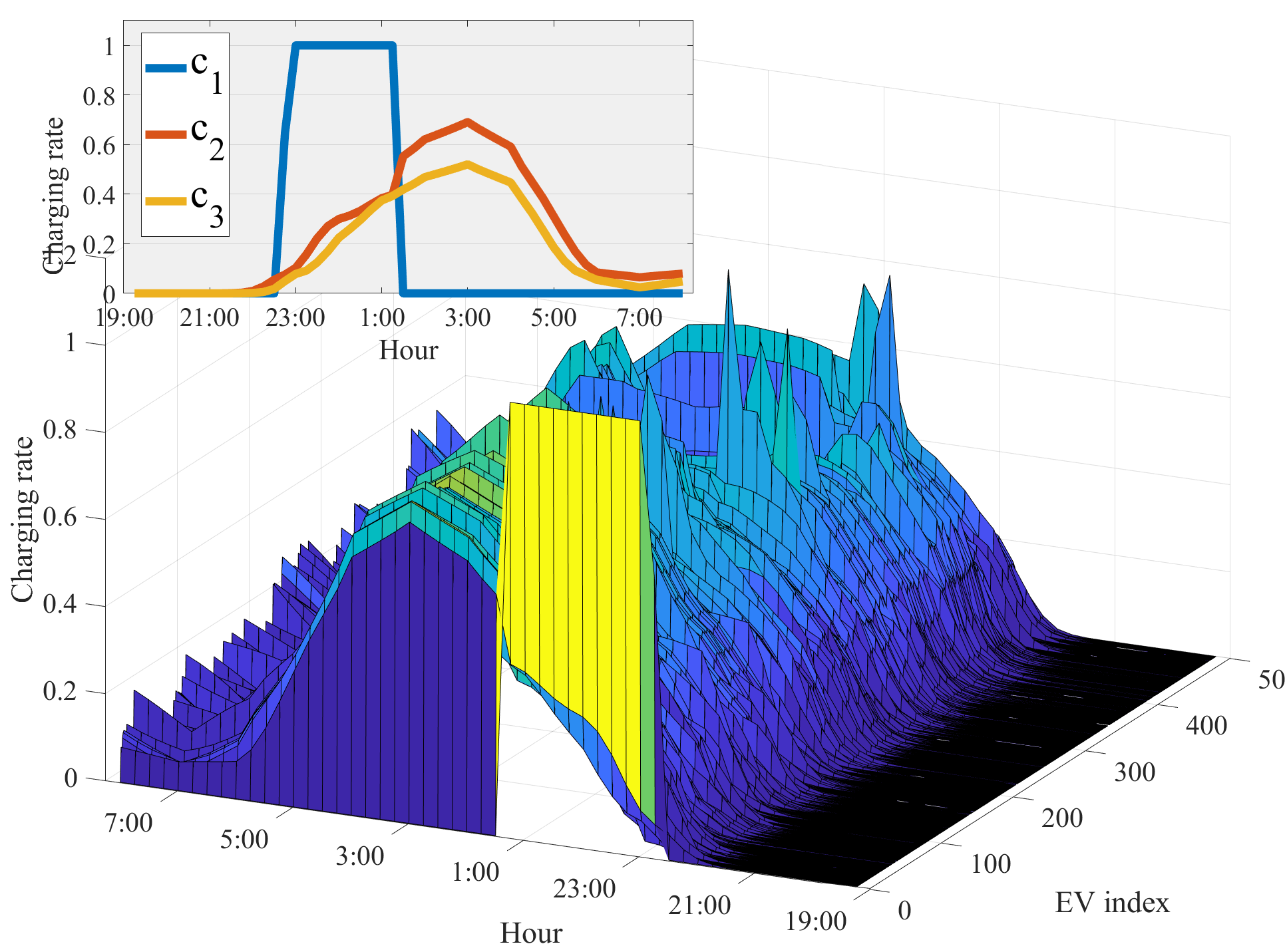}\label{prime desired u}}\\[0.75ex]

  (b){\includegraphics[width=0.30\textwidth]{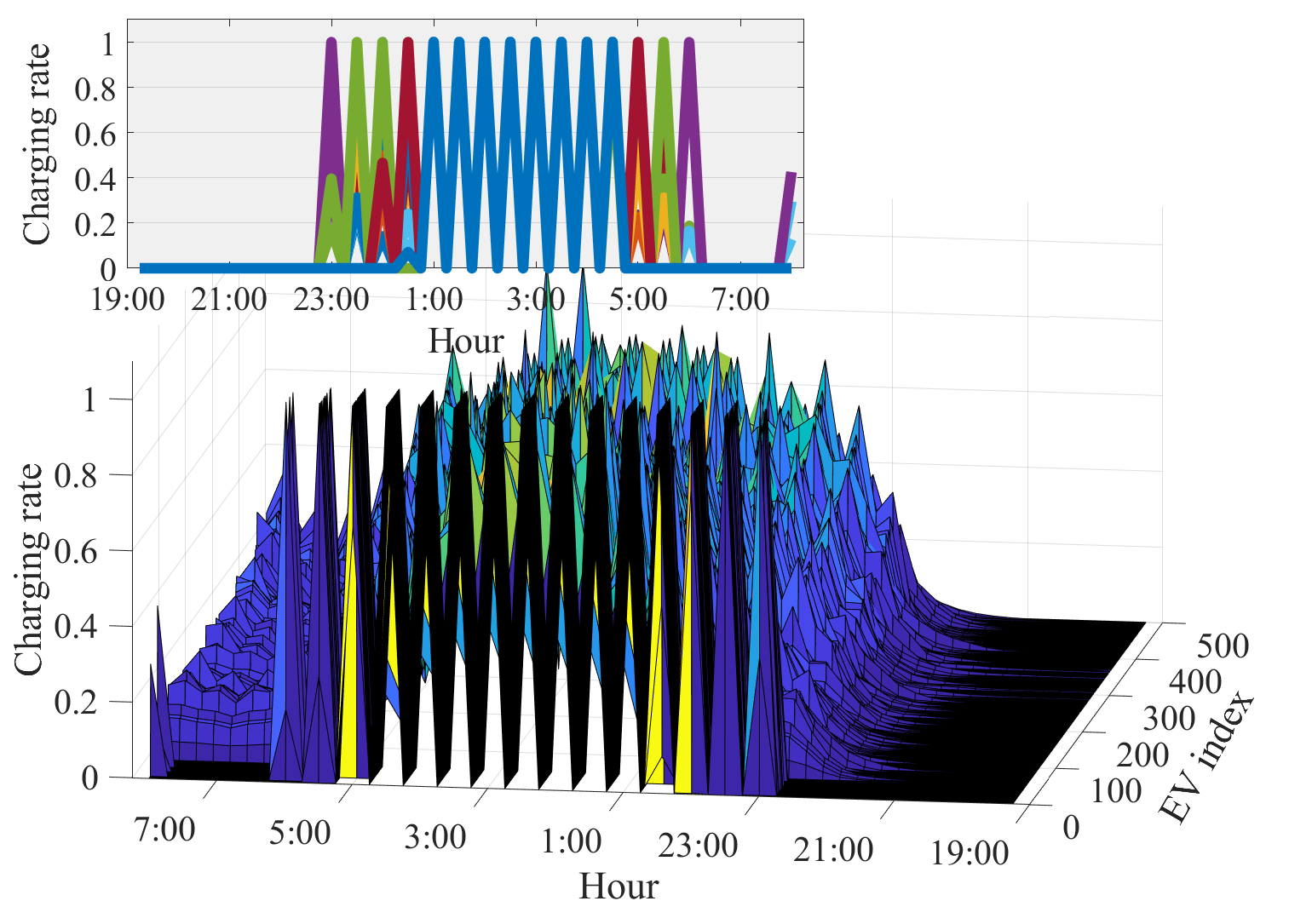}\label{battery damage attack u}}\hfill
  (d){\includegraphics[width=0.30\textwidth]{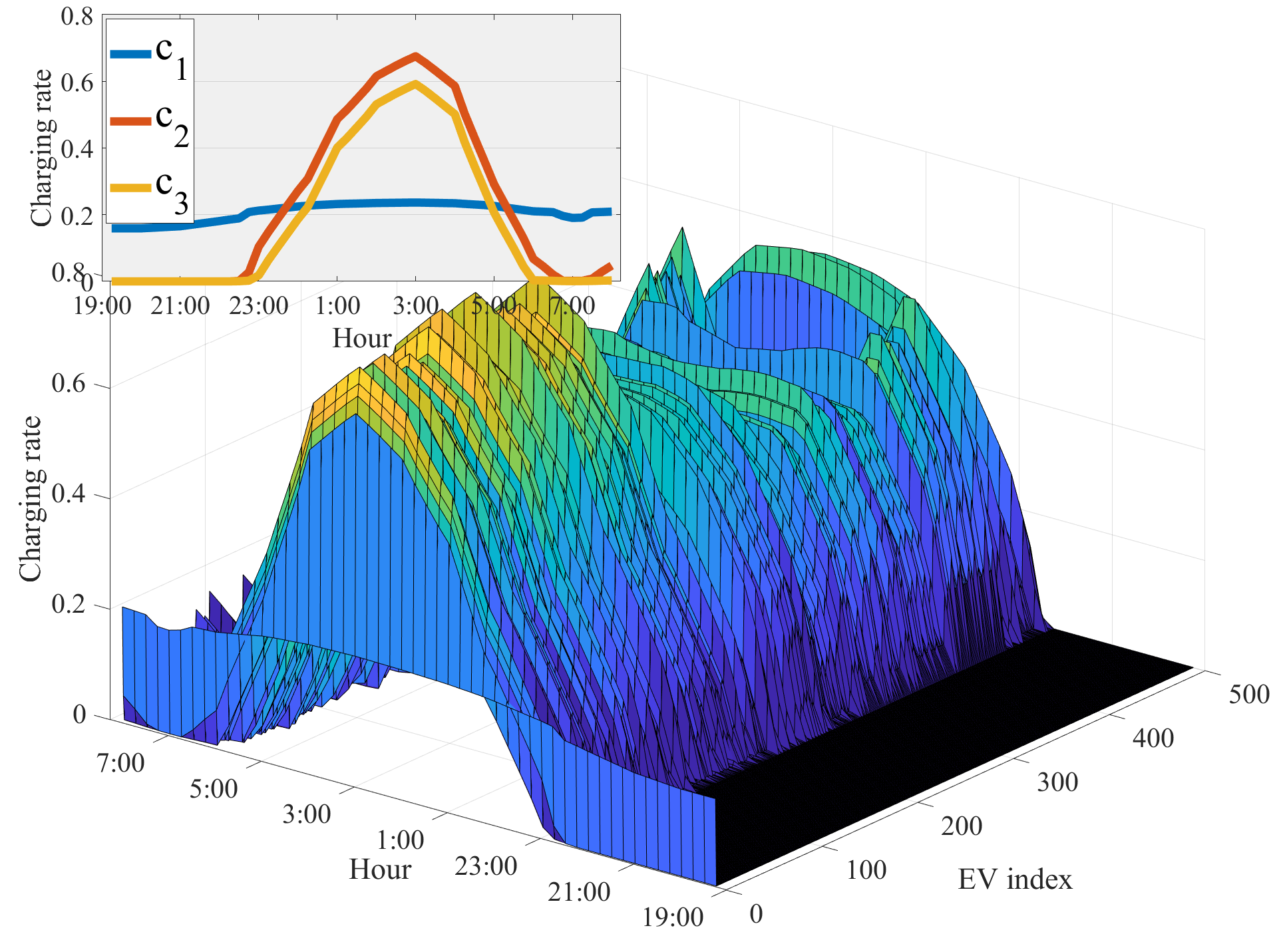}\label{dual attack u}}\hfill
  (f){\includegraphics[width=0.30\textwidth]{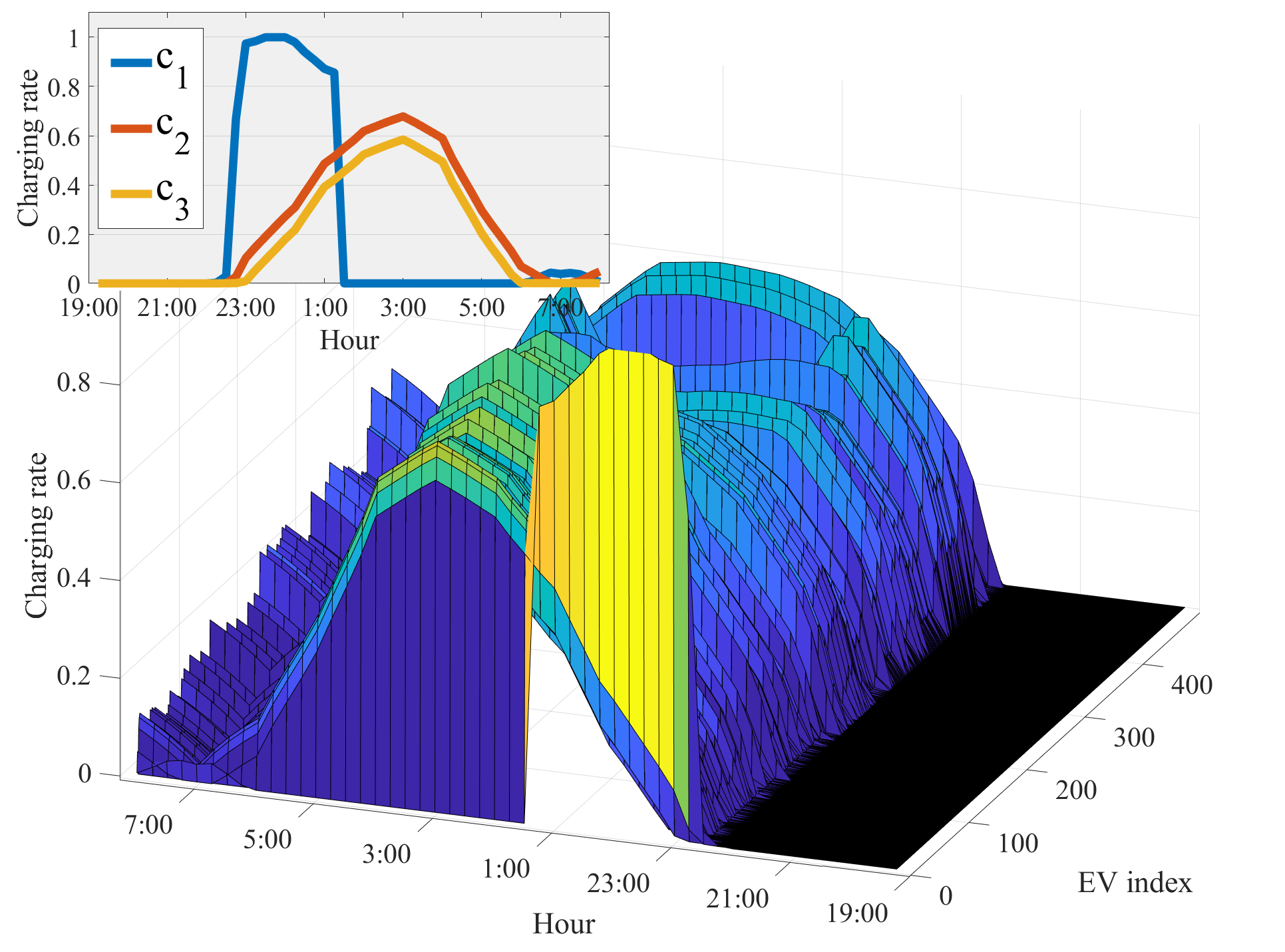}}\label{dual desired u}
  \caption{Charging profiles of all EVs under (a) No attack, (b) Primal battery damage, (c) Primal smooth-charging, (d) Dual smooth-charging, (e) Primal time-tuning, and (f) Dual time-tuning attack.}
  \label{all}
\end{figure*}

\begin{figure}[!htb]
    \centering  
\includegraphics[width=0.35\textwidth, trim={0cm 0.4cm 0.0cm 0.5cm},clip]{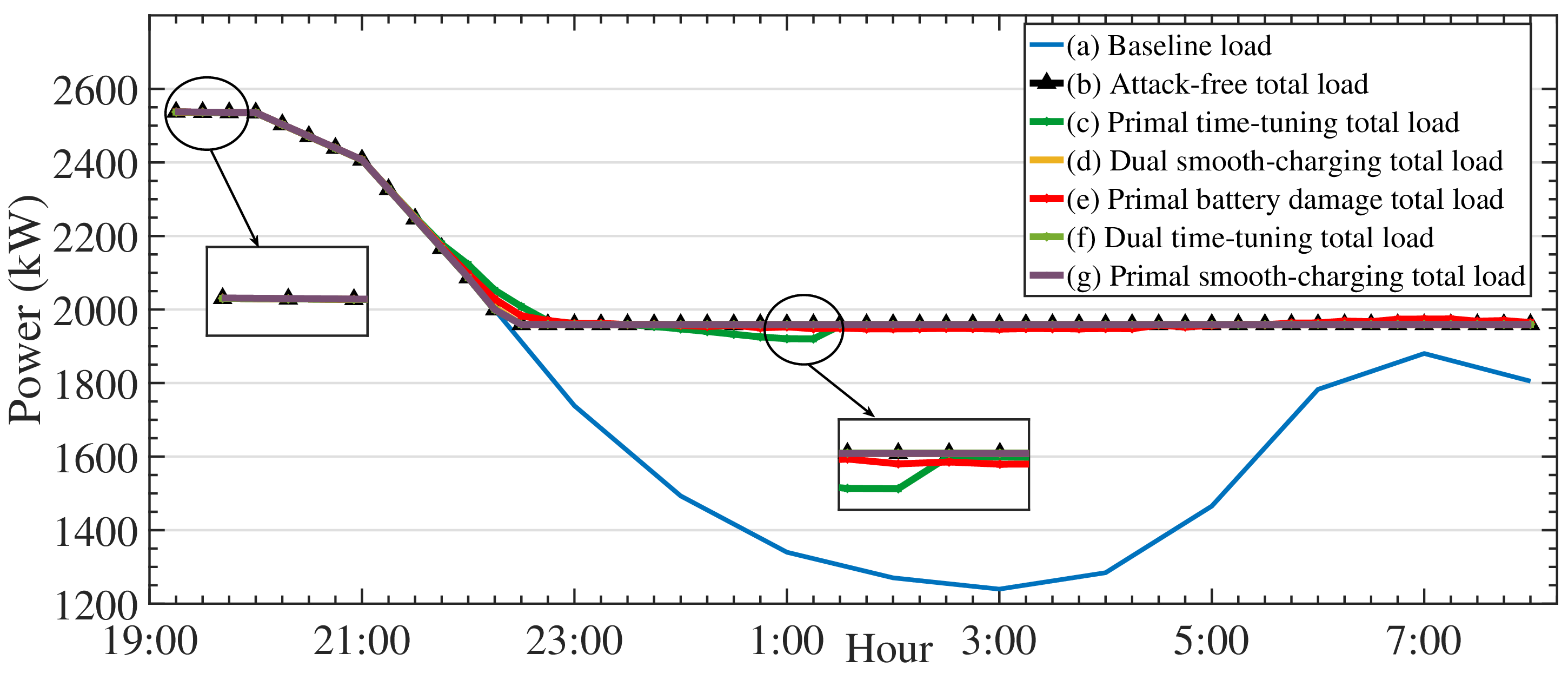}
    \caption{Baseline load and total load under different attacks.}
    \label{load profiles}
\end{figure}

\begin{figure}[!htb]\centering\includegraphics[width=0.485\textwidth, trim={0cm 0.8cm 0.0cm 0.4cm},clip]{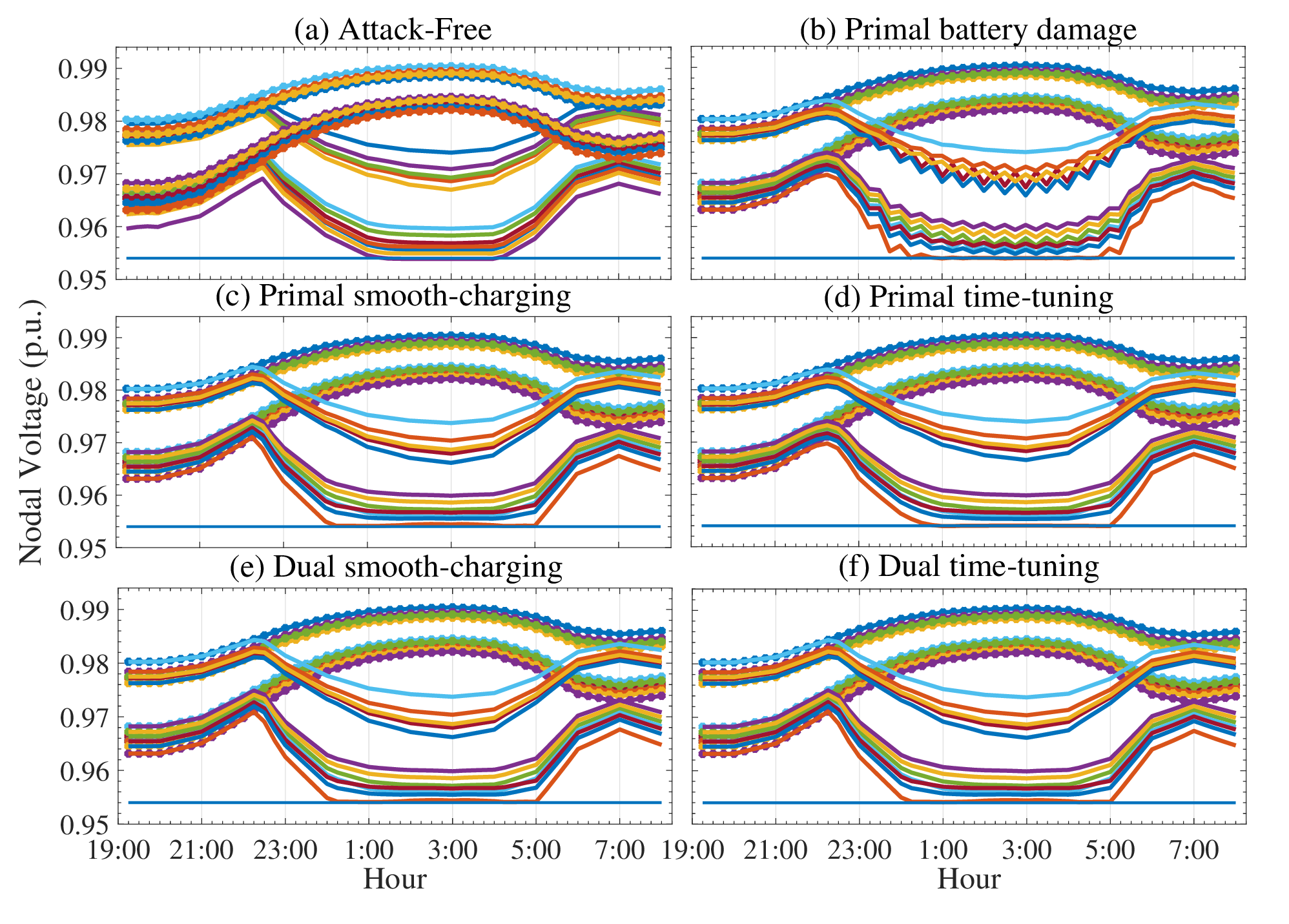}
\vspace*{-0.5cm}
    \caption{Nodal voltage magnitudes. Star-marked lines represent the baseline case, and solid lines represent the case with controlled EV charging loads.}
    \label{v_compact}
\end{figure}

By running SPDS to solve the attack-free problem in \eqref{main problem}, the charging profiles of all 500 EVs are shown in Fig. \ref{all}(a).
The converged valley filling performance and the nodal voltage magnitudes under the attack-free scenario are shown in Fig. \ref{load profiles}(b) and  Fig. \ref{v_compact}(a), respectively.   It can be observed that the total load profile becomes flat and stays at $1,958$ kW after 22:30, and all nodal voltage magnitudes are maintained above 0.954 p.u. 

\subsubsection{Battery damage attack}
Let the first 50 EVs be attackers (victims). The attack power is set to $\omega_1=1$. $m$ and $M$ are selected as $0.2$ and $1\times10^5$, respectively. $t_f$ is selected as $2$, indicating that the charging rate is expected to oscillate every time step, causing more damage to batteries. Fig. \ref{all}(b) shows that the charging profiles of the victim EVs severely oscillate between zero and one. These high-frequency, large-amplitude rate changes are detrimental to battery state of health. Moreover, all other EVs are experiencing spikes of the same frequency in their charging profiles due to the oscillatory charging of the first 50 EVs.  
Fig. \ref{v_compact}(b) presents the nodal voltage magnitudes under this attack. It shows that this attack not only affects bus 2, where the victim EVs are connected, but also creates oscillations in other nodal voltage profiles.  Though all nodal voltage magnitudes are still above the lower bound and follow a similar pattern, the oscillation is highly noticeable and can alert the system operator to an anomaly.

\subsection{Dual attack scenarios}
\begin{figure}[!htb]
    \centering
{
        \includegraphics[width=0.48\columnwidth]{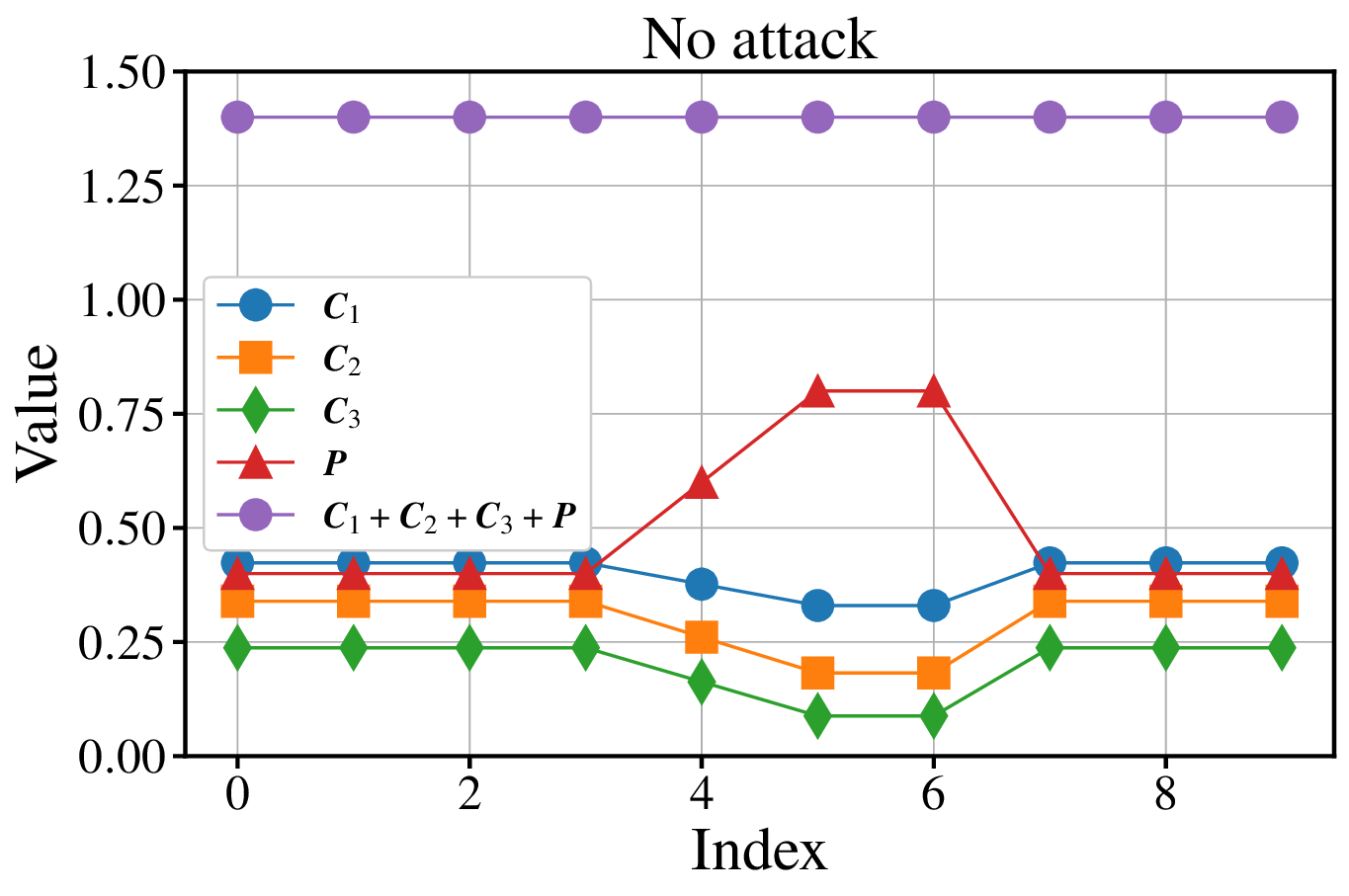}%
        \label{fig:case1}
    }
    \hfill
{
        \includegraphics[width=0.48\columnwidth]{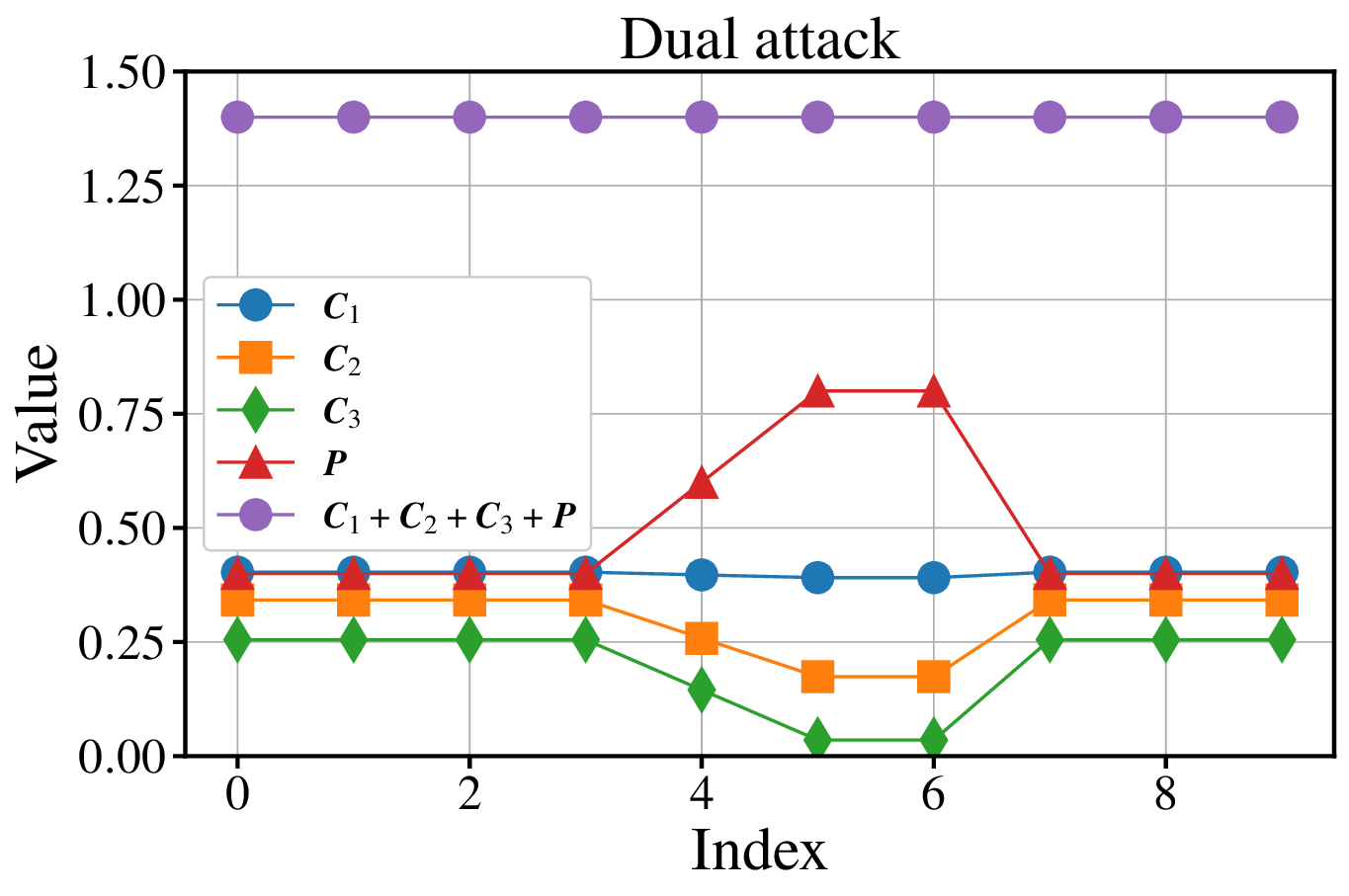}%
        \label{fig:case2}
    }
{
        \includegraphics[width=0.48\columnwidth]{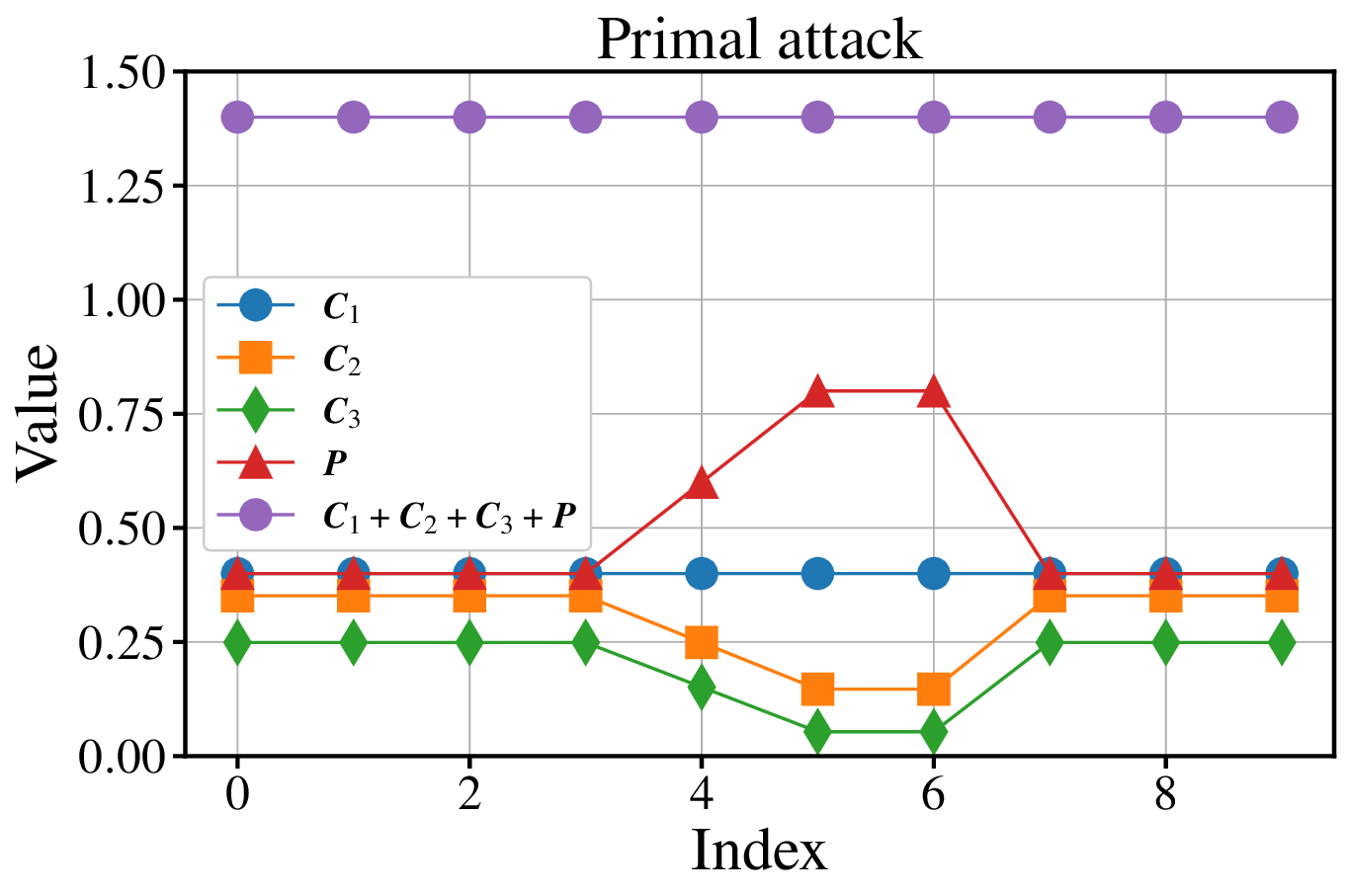}%
        \label{fig:case3}
    }
    \hfill
    {
        \includegraphics[width=0.48\columnwidth]{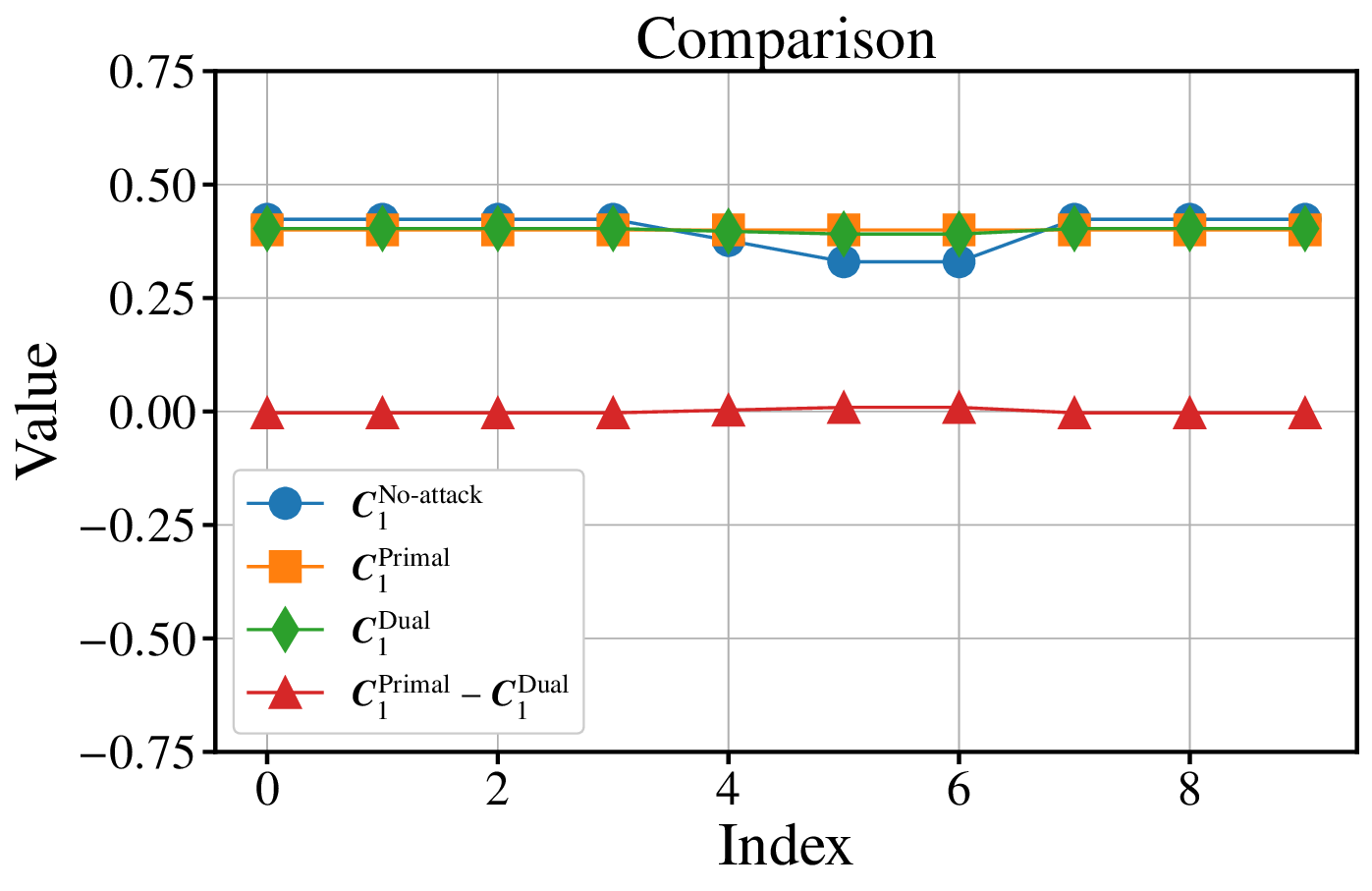}%
        \label{fig:case4}
    }
    \vspace*{-0.8cm}
    \caption{Charging profiles of the 3-EV toy problem under dual attack}
    \label{toy problem}
\end{figure}

We first present a toy problem to show the effectiveness of dual attacks before implementing it in EV charging control. Herein, we consider the original problem in \eqref{main problem} with only three EVs and replace \eqref{3c} and \eqref{3b} with $\bm{C}_1 + 0.2  \bm{C}_2 + 0.3  \bm{C}_3 \leq \bm{b}$ and $\|\bm{C}_i\|_2={\hat{b}}_i$, respectively, where $\bm{b}$ is a constant vector and ${\hat{b}}_i$'s are scalars. Let EV 1 be the attacker pursuing a smooth charging profile. In the primal attack, EV 1 injects $2\bm{C}_1$ into its own channel, while in the dual attack case, it manipulates the other two EVs' communication to inject the gradient of $\|\bar{\bm{c}}- 0.2 \bm{C}_2 - 0.3\bm{C}_3\|_2^2$ into their communication channels. The challenging part for the attacker in the dual case is to find the proper $\bar{\bm{c}}$, which comes from the constraint $\bm{C}_1 + 0.2 \bm{C}_2 + 0.3 \bm{C}_3 \leq \bm{b}$ and the estimated value of $\bar{\bm{c}}=\bm{C}_1^\star + 0.2 \bm{C}_2^\star + 0.3 \bm{C}_3^\star$ where $\bm{C}_i^\star$ are the optimal values from the primal attack case. The similarity of the results between primal and dual attacks depends on how accurately the attacker can forecast the optimal values from the primal attack to design its dual attack vector. The toy problem results are shown in Fig. \ref{toy problem}, where it can be seen that the attacker achieved its goal in both attack cases by flattening its charging profile.

In the following dual attack scenarios with  500 EVs, we consider EV 1 as the attacker who is connected to bus 2 along with EV 2. For simplicity, the attacker is set to only exploit EV 2 to deploy the dual attack. However, the attacker can manipulate more victims to achieve its purpose. We implemented {Lemma 2} in the following dual attack scenarios. 
\subsubsection{Dual smooth-charging attack}
\begin{figure}[!htb]
    \centering
    \includegraphics[width=0.8\linewidth, trim={0cm 0.4cm 0.0cm 0.2cm},clip]{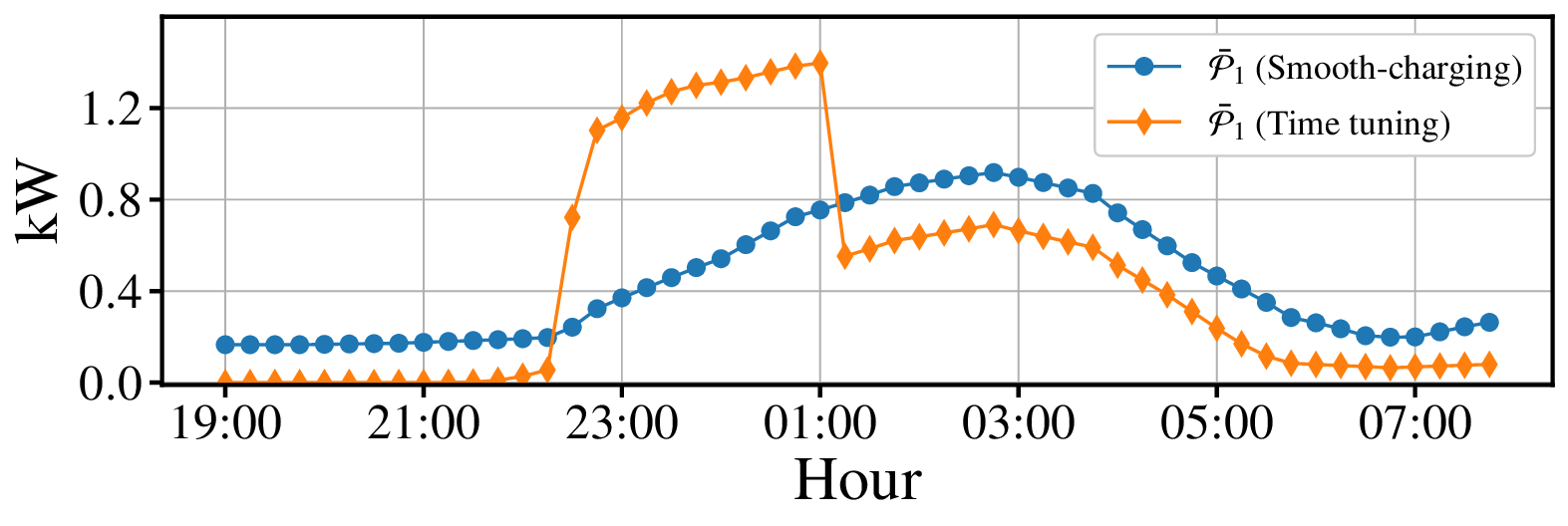}
    \vspace*{-0.2cm}
    \caption{Values of calculated vectors $\bm{\bar{\mathcal{P}}}_1$ for dual attack scenarios}
    \label{p1}
\end{figure}

The nodal load balance at bus 2 can be written as $\bm{c}_1=\bm{\bar{\mathcal{P}}}_1-\bm{\bar{\mathcal{P}}}_{21}\bm{c}_{2}$. For a purely smooth charging profile, EV 1 needs a constant $\bm{c}_1$ to satisfy $E_{1,req}-\bm{\hat{B}}_{1,2}\bm{c}_1=0$. The attacker knows $E_{1,req}=7.72\times 10^3$ and $B_1=-675$, and it calculates the desired charging rate for a smooth-charging attack, $\hat{c}_1(\hat{t})=0.2,~ \forall\hat{t}\in T$. Further, if the attacker can successfully maintain the power balance equation with EV 2, the charging profile of EV 2 will become dependent on that of EV 1. Moreover, EV 1 can manipulate the information of $\bm{\bar{\mathcal{P}}}_1$ in such a way that the resulting $\bm{c}_1$ has the desired $\hat{c}_1(\hat{t})$. For this case, we calculated  $\bm{\bar{\mathcal{P}}}_1$ as shown in Fig. \ref{p1}  to achieve smooth-charging for EV 1.
Fig. \ref{all}(c) shows the charging profiles of all EVs when EV 1 is deploying the primal smooth-charging attack. It can be seen that the charging rate of EV 1 follows the desired charging rate, with a mean of $0.205$ and a standard deviation of $1.484$. In the meantime, $\bm{c}_2$ has a mean of $0.246$ and a standard deviation of $2.551$.

The charging profiles of all EVs when EV 1 launches the dual attack are shown in  Fig. \ref{all}(d). In this case, the charging profile of EV 1 presents a similar pattern as in the primal attack case with a mean of $0.208$ and a standard deviation of $1.5108$. This indicates that the dual attack can successfully achieve a smooth charging profile with a negligible $1.5\%$ mean value deviation from the primal attack case. Meanwhile, the charging profile of EV 2 also presents subtle changes of about $2\%$ two-norm percentage change from the primal attack case.


\subsubsection{Dual time tuning attack}

With the same configuration presented in the previous scenario, EV 1 is set to get its battery fully charged as soon as possible, starting from $T=16$. Fig. \ref{all}(e) shows the ideal charging profile of EV 1 under the primal attack.     
In the dual attack case, EV 1 ideally intends to establish the same charging profile. To this end, the desired $\bm{\bar{\mathcal{P}}}_1$ is calculated as shown in Fig. \ref{p1}. 
Fig. \ref{all}(f) shows all EVs' charging profiles under the dual time tuning attack by EV 1. Notably, the time tuning goal of EV 1 has been successfully achieved, though with minor deviations from the ideal curve. In the primal attack case, EV 1 maintains its charging rate at $1.0$ between $T=16$ and $T=25$. In contrast, in the dual attack case, the charging rate remains at $1.0$ from $T=16$ through $T=20$, then gradually drops to $0.85$ at $T=25$. 


\section{Conclusion} \label{sec V conclusion}
This paper developed a principled foundation for targeted algorithmic cyber-attacks in distributed multi-agent optimization (DMAO), introducing a unified framework that spans primal and dual cyber-attacks, and demonstrating its practicality on distributed EV charging control. We formalized the attack design under standard convexity and feasibility assumptions, derived sharp deviation guarantees that bound the distance between attacked and nominal optima via solid theorems, and established an exact equivalence between dual and prime attacks. These results show that structurally aligned, stealth-preserving manipulations can materially reshape system outcomes while maintaining algorithmic convergence and constraint satisfaction. Simulations on large-scale EV fleets corroborate the theory, revealing measurable shifts in charging trajectories and system metrics and illustrating the trade-off between attacker benefit and stealthiness. Looking forward, we will work on tightening the bounds derived here to suggest defense directions that minimize worst-case deviation. Further, we will work on detection methods to accurately identify algorithmic attacks and explore the possible mitigation techniques.

\bibliographystyle{IEEEtran}
\bibliography{Refrences}

\end{document}